\documentclass[useAMS,usenatbib]{mnras}

\usepackage{graphicx,natbib,url,twoopt}

\usepackage{amsmath}
\usepackage{amssymb}

\bibpunct{(}{)}{;}{a}{}{,}    
\makeatletter
 \newcommandtwoopt{\citeads}[3][][]{\href{http://adsabs.harvard.edu/abs/#3}%
   {\def\hyper@linkstart##1##2{}%
    \let\hyper@linkend\@empty\citealp[#1][#2]{#3}}}    
 \newcommandtwoopt{\citepads}[3][][]{\href{http://adsabs.harvard.edu/abs/#3}%
   {\def\hyper@linkstart##1##2{}%
    \let\hyper@linkend\@empty\citep[#1][#2]{#3}}}      
 \newcommandtwoopt{\citetads}[3][][]{\href{http://adsabs.harvard.edu/abs/#3}%
   {\def\hyper@linkstart##1##2{}%
    \let\hyper@linkend\@empty\citet[#1][#2]{#3}}}      
 \newcommandtwoopt{\citeyearads}[3][][]%
   {\href{http://adsabs.harvard.edu/abs/#3}%
   {\def\hyper@linkstart##1##2{}%
    \let\hyper@linkend\@empty\citeyear[#1][#2]{#3}}}   
\makeatother

\newcommand{\rem}[1]{ } 

\def\lcdm{$\Lambda$CDM\ }

\title[Dark matter halos in the 2cDM model]
  {Dark matter halos in the multicomponent model. I. Substructure}
\author[]
  {Keita Todoroki,$^1$\thanks{Email: keita@ku.edu}
  Mikhail V. Medvedev$^{1,2}$
  \newauthor 
             \\
  $^1$Department of Physics and Astronomy, University of Kansas, Lawrence, KS 66045\\
  $^2$Laboratory for Nuclear Science, Massachusetts Institute of Technology, Cambridge, MA 02139
  }
\date{\today}

\pagerange{\pageref{firstpage}--\pageref{lastpage}} \pubyear{2016}

\def\LaTeX{L\kern-.36em\raise.3ex\hbox{a}\kern-.15em
    T\kern-.1667em\lower.7ex\hbox{E}\kern-.125emX}

\begin{document}


\label{firstpage}

\maketitle

\begin{abstract}
Multicomponent dark matter with self-interactions, which allows for inter-conversions of species into one another, is a promising paradigm that is known to successfully and simultaneously resolve major problems of the conventional $\Lambda$CDM cosmology at galactic and sub-galactic scales. In this paper, we present $N$-body simulations of the simplest two-component (2cDM) model aimed at studying the distribution of dark matter halos with masses $M\lesssim10^{12}M_\odot$. In particular, we investigate how the maximum circular velocity function of the halos is affected by the velocity dependence of the self-interaction cross-sections, $\sigma(v)\propto v^a$, and compare them with available observational data. The results demonstrate that the 2cDM paradigm with the range of self-interaction cross-section per particle mass (evaluated at $v=100$~km~s$^{-1}$) of $0.01\lesssim \sigma_0/m\lesssim 1$~cm$^2$g$^{-1}$ and the mass degeneracy $\Delta m/m\sim 10^{-7}-10^{-8}$ is robustly resolving the substructure and too-big-to-fail problems by suppressing the substructure having small maximum circular velocities, $V_{\rm max}\lesssim100$~km~s$^{-1}$. We also discuss the disagreement between the radial distribution of dwarfs in a host halo observed in the Local Group and simulated with CDM. This can be considered as one more small-scale problem of CDM. We demonstrate that such a disagreement is alleviated in 2cDM. Finally, the computed matter power-spectra of the 2cDM structure indicate the model's consistency with the existing Ly-$\alpha$ forest constraints.
\end{abstract}



\section{Introduction}

Dark matter (DM) is believed to be one of the most crucial constituents of the Universe that shapes its  overall structure and cosmological history. Gravity of the DM cosmic web largely determines the dynamics of baryons on large scales, where and how the galaxies form and evolve. 
Over the past decades, the collisionless Cold Dark Matter (CDM) paradigm with a cosmological term ($\Lambda$CDM) has become the most successful cosmological model that is able to reproduce the nonlinear evolution of the large-scale structure of the observed Universe. However, there is growing observational evidence that the conventional \lcdm model may be inaccurate at small scales. There are several ``small-scale" problems as follows. 

One of the long-standing problems is the `missing satellites' problem or the `substructure' (SS) problem \citep{klypin1999,moore1999}. Namely, numerical simulations of \lcdm systematically over-predict the number of small-mass (satellite) halos compared to observations in the Milky Way (MW) environment \citep[see, e.g.,][]{diemand2008}. The problem also extends out to a much larger volume of space, in which the numerically predicted number of galaxies in the field is much higher than observed \citep{zavala2009,papastergis2011,klypin2015}.
In the past decade, numerous ultra-faint dwarf galaxies have been discovered in the proximity of the MW \citep[e.g.,][]{willman2005,belokurov2007,bechtol2015,laevens2015}, which has increased the number of observed substructures in the local environment roughly by factor of a few, hence the discrepancy with $\Lambda$CDM predictions has been reduced at the lower-mass end. 
In addition, our MW's closest neighboring galaxy M31 has been known to host a similar population and distribution of satellites, although \citet{richardson2011} found the brighter M31 dSphs tend to have larger half light radii than their MW counterparts. It is possible that there remains a yet undetected population of faint dwarf galaxies ($M_{\star} \lesssim 10^{5}M_{\odot}$), which would further reduce the discrepancy. However, such a population is not expected to be large, given that the faintest (with the mass-to-light ratio of about a few hundred) MW satellites detected have the stellar mass of only about $10^{3}M_\odot$ \citep{simon2007,simon2011}.

Several possible resolutions to this problem have been proposed. The simplest solution, in which internal heating mechanism can destroy host's satellites by stellar feedback originating from the residing dwarf galaxies, has been shown to be unlikely \citep{penarrubia2012}. In an alternative scenario, external heating source from the photoionizing/UV background could set a critical halo mass below which baryonic material cannot remain trapped in the host halos, thus reducing the substructure \citep{efstathiou1992, quinn1996, summerville2002, okamoto2008, hambrick2011}. Whether this scenario applies to highly DM-dominated halos remains unclear. Tidal stripping can also be a mechanism to disrupt substructure \citep{hayashi2003}, but see \cite{kazantzidis2004}.
In general, whereas baryons can have strong effect in large halos, their role in ultra-faint, DM-dominated, low-mass dwarfs cannot be large. 

The second, somewhat related, problem of the CDM model is the so-called `too-big-to-fail' (TBTF) problem \citep{boylan-kolchin2011}. 
It was found that MW-type environments in numerical simulations contain a larger number of massive subhalos with denser halo center than what is observed, and they should be able to host observable galaxies, if they actually exist.  
In a sense TBTF problem is effectively the substructure problem at the higher mass end \citep{papastergis2015}. A possible solution to the TBTF problem is based on the baryonic physics, namely the stellar feedback, analogous to the generally favored solutions for the SS and core-cusp (discussed below) problems. However, \cite{garrison-kimmel2013} showed that the supernova feedback alone is not capable of solving the TBTF problem, even if the feedback is strong. Observationally, \cite{papastergis2016} and \cite{papastergis2015} found that the baryonic effects, including reionization feedback and taking into account the rotation curves of halos, do not seem to resolve the problem either. 
Studies by \cite{brook2015} showed, however, that the problem can be alleviated by choosing a mass-dependent density profile, which seems a rather {\em ad hoc} solution.

The third problem of the CDM model is the `cusp-core' (CC) problem, in which DM-only \lcdm simulations predict cuspy, $\rho\propto 1/r$, halo density profiles \citep{flores1994, navarro1996, klypin2001}. This contrasts with flattened cores observed in centers of dark matter-dominated systems, such as dwarfs, dwarf spheroidals (dSph) and low surface brightness (LSB) galaxies \citep{deblok2001, deblok2002}. Interestingly, the cuspiness of the inner profile can even be enhanced by the baryonic infall to the DM halo, which is implied both theoretically \citep{blumenthal1986} and observationally \citep{swaters2009}. 
Recently, \citet{errani2017} reported that a cuspy MW-type halo profile produces a larger number of substructures than a cored one, which implies a strong connection between the CC and SS problems. A plausible solution to the cusp-core problem is the stellar feedback that is energetic enough to disrupt or alter the halo core by providing thermal energy to slow down baryon accretion and by transferring kinetic energy to DM so that the inner density is reduced, transforming the inner profile from a cuspy to a flatter, i.e., cored one \citep{pontzen2012, governato2012}.
This, however, requires a repeated bursty star formation (SF) processes in the center of galactic halo with a sufficiently high efficiency in energy transfer to DM. 
Whether such a mechanism can be achieved in DM-dominated dSphs, for example, is questionable. 
Some authors argue that those systems used to have active star formation do create cored profiles at earlier times, but after depleting the stellar reservoir and continuously accreting mass at later times the slope of the inner profile could have deviated from the nearly flat one to a steeper one \citep{kormendy2016}. Similarly, \cite{Onorbe2015} found that only those simulated dwarf galaxies that experienced late-time SF are able to create DM cores. They argued that early-time SF only temporarily flattens the inner halo profile, and if the SF does not persist until later time, the once-cored halo is turned into a cuspier, CDM-like halo. 
These results might indeed explain why some observations show higher inner densities in dSphs compared to the larger systems.
The cored density profile, although not nearly as cored as that of some dwarfs, can also be seen in observed galaxy clusters \citep{newman2013b, collett2017}. 
For such large systems, numerical studies show that strong AGN feedback from Super Massive Black Holes is able to create a cored inner profile \cite{martizzi2012, martizzi2013}. 
As such, the presence of the non-cuspy density profiles appears to be ubiquitous across many orders of magnitude in mass scale. Strong baryonic feedback processes alone do not seem to resolve the CC problem over such a wide mass range, and it is particularly so for dark matter dominated systems.

To summarize, solving all the above three major cosmological problems of the CDM model simultaneously and consistently appears to be increasingly challenging, as it requires {\em ad hoc} or fine-tuning of the baryonic physics models, especially the `realistic' stellar feedback which, until now, is still implemented as subgrid prescriptions due to the lack of numerical capabilities. Although numerous baryonic feedback models have been created thus far, the detailed physics that governs the whole process remains largely uncertain. The question one might ask is: What if we will be able to adjust available and rather flexible feedback models so that the simulations fully reproduce the observable Universe? Is this the much-needed resolution of the problems? We do not think that this is the best approach, because viable subgrid baryonic models must be physically valid. Yet, the validity of them is hard to justify with the existing computational resources. Given the fact that much fine-tuning is already needed to resolve the problems, perhaps a simpler solution would be to re-assess our assumptions about DM physics. Thus, alternative DM models aimed at solving the small-scale problems have already been proposed. The most popular alternative DM models are the Self-Interacting Dark Matter (SIDM) model, the Warm Dark Matter (WDM) model and the Fuzzy Dark Matter (FDM).

SIDM was proposed by \cite{spergel2000} as a promising solution to the dense DM cusps (i.e. the CC problem, but not the SS and TBTF problems), while keeping CDM predictions unchanged at large scales. It essentially allows DM particles to scatter off each other in addition to gravitational interactions. $N$-body numerical simulations fully confirmed this idea, provided the interaction cross-section is within the acceptable range. That is, it needs to be not too large to avoid gravothermal catastrophe which creates very cuspy profiles $\rho\propto r^{-2}$, and not too small to make a sizable effect over the Hubble time \citep[e.g.][]{yoshida2000, moore2000, rocha2013}. 
For the inferred values of the cross-section per unit mass $\sigma/m$, theoretical works suggest 0.1 to 10 cm$^{2}$g$^{-1}$ \citep[e.g.][]{randall2008, rocha2013, vogelsberger2013}, whereas observations generally imply $\sigma/m$ $\lesssim$ 1 $\rm cm^{2}g^{-1}$, which comes, e.g., from the numerical modeling of the Bullet Cluster \citep{markevitch2004}. Recently, \citet{kamada+17} have shown that the SIDM model with $\sigma/m=3$~cm$^{2}$g$^{-1}$ provides excellent fits to the rotation curves of galaxies with asymptotic velocities being in the 25--300~km~s$^{-1}$ range.

In general, SIDM does solve the CC problem, but it absolutely cannot solve the SS and TBTF problems. This is because scattering between DM particles can only transfer energy (temperature) to make the halo nearly isothermal in the inner halo, but they cannot destroy small halos or else appreciably reduce their mass. Thus, the cumulative number of halos in the lower mass range predicted by the numerical simulations with the SIDM model is nearly identical with that of CDM and thus at odds with observations. In particular, the TBTF and SS problems state that there is a characteristic velocity ($\sim50-100$~km~s$^{-1}$) and hence the mass ($\sim10^{10}M_\odot$) of the halos, below which the halos are `missing' in the observed MW-type environment. These characteristic values do not appear in a simple elastic scattering mechanism offered by SIDM.  

In principle, maxwellianization of the particle distribution function by collisions creates an exponential tail of escaping particles, which could lead to halo evaporation. However, this process is very slow and requires a large number of collisions, which would lead to gravothermal collapse and the formation of even steeper $\rho\propto r^{-2}$ cusp, which is not seen observationally. The collapse thus limits the number of particle interactions to be a few, at most, over the Hubble time. In this case, thermal evaporation in SIDM is negligible.  Thus, we note that regardless of a  velocity-dependent cross-section model at hand, SIDM model cannot resolve the SS and TBTF problems.
In the meantime, SIDM with baryons has also been studied recently by \cite{fry2015}, whose study implies that SIDM with a velocity-dependent cross-section would be required if the faint dwarfs are to have realistic DM cores. Recent simulations with baryonic physics \citep{vogelsberger2014} indicate that SIDM does not significantly alter the stellar concentration and distribution, even though the stellar mass distribution is slightly expanded with a reduced density at the central region of $< 1$~kpc.

Next, WDM was once considered to be the ultimate solution to all the problems because finite thermal velocities of DM introduce a natural scale --- the free streaming length --- below which density fluctuations are exponentially suppressed. WDM does suppress both the mass function and the density profile cusps below this scale \cite[see, e.g.,][]{zavala2009}. This indeed works well for MW-type galaxies and their large satellites but fails for small dwarf galaxies, whose observed cores are too small to be consistent with the WDM predictions. Apparently, a single scale is not enough. Different scales are needed to reconcile the dwarf profiles and the mass function. Thus, WDM is not capable of solving the SS and TBTF problems \citep{klypin2015, papastergis2015}. 

The FDM model assumes that a dark matter particle has a very small mass, $\sim 10^{-22}$~eV, so that its de Broglie wavelength, $\lambda_{dB}$, is of the order of a kpc \citep{turner1983, hu2000, hui2017}. This way, the presence of cores is naturally explained as $r_c\sim\lambda_{dB}$. The substructure problem is explained by tunneling of small halos in the tidal potential of a larger parent halo. However, this model faces some problems with the Ly$\alpha$-forest constraint \citep{hui2017}. Furthermore, the tunneling probability is an exponential function of the distance, so the model would predict very rapid disappearance of dwarf halos inside a certain critical radius and nearly intact dwarf population outside of it. Such a distribution of halos with the distance from the parent halo center seems to be at odds with observations.

In summary, it seems rather difficult to resolve the CC, SS and TBTF problems simultaneously. Therefore, more sophisticated multi-species DM models have been proposed \citep{graham2010,mccullough2013,bramante2016,kuflik2016}.
Recently, \cite{vogelsberger2016} studied the effects of introducing DM interactions with a massless neutrino-like fermion (dark radiation) in $N$-body simulations. Their results showed that the internal structure and the abundance of subhalos are greatly affected by both self-interactions and the small scale primordial damping of the power spectrum with a velocity-dependent cross-section, implying a possibility of solving or alleviating the SS and TBTF problems.

Among alternative ideas, the multicomponent flavor-mixed DM model  ($N$cDM),  or just a 2-component model in its simplest incarnation (2cDM), is particularly interesting because it can resolve all the problems simultaneously, yet it does not violate all known constraints \citep{medvedev2010, medvedev2010b, medvedev2014theo, medvedev2014}. We should note here that our numerical simulations do not explicitly simulate quantum effects --- this is a formidable task for cosmological simulations with present-day computing resources. Instead, we model interactions of mass-eigenstates simply as interactions of DM particles of various types. Thus, our simulations effectively represent all models with several DM species that are allowed to inelastically and elastically interact with each other. However, our flavor-mixed model is the only one, to our knowledge, that successfully and naturally evades the early universe constraint \citep{medvedev2014theo}. The latter is formulated as follows. If a multicomponent model has self-interactions which substantially reduce the abundance of `excited' or `heavier' species at the present epoch (otherwise there would be no effect on halos), the very same process should also happen in the early universe when the DM density is orders of magnitude higher. Thus, all the `excited' or `heavier' species must be completely destroyed (i.e., exponentially suppressed by the Boltzmann factor) at a very high $z$, so that the model effectively becomes a standard single-component CDM or SIDM. Thus, every viable multicomponent self-interacting DM model should provide a mechanism to avoid this early universe constraint. 

Hereafter, we consider the two-component dark matter (2cDM) model for simplicity, which is comprised of  `heavy' and `light' mass eigenstates with masses $m_h>m_l$. The model assumes, in analogy with SIDM, interactions between DM particles. Their flavor-mixed nature results in a non-zero probability of inter-conversions of the DM-particle mass eigenstates \citep{medvedev2010b, medvedev2014theo}. As in SIDM, the DM particles can scatter and create halo cores. Additionally, the low-mass halos with masses below a certain threshold, $M_0$, can lose their masses via `quantum evaporation' or the `Baron M\"unchausen effect'\footnote{This name is after Baron von M\"unchausen, a character of ``The Surprising Adventures of Baron M\"unchausen'' by R.E. Raspe, who, in one of his ``true'' stories, lifted himself and his horse out of the mud by pulling on his own pigtail.}. 
Correspondingly, this process flattens the slope of the mass function below $M_0$. This `evaporation' occurs because in mass-eigenstate conversions, the energy $\Delta mc^2 \equiv (m_{h} - m_{l})c^2$ is released in the form of kinetic energy that is enough to kick a DM particle out of the halo. The value of $M_0$ depends on the fractional difference in the masses of the species, $\Delta m/m$, whereas the amplitude of the mass-function suppression and the core density are related to the interaction cross-section, which we assume to be velocity dependent, $\sigma(v)\propto v^a$ (with $a$ being constant). We assume that $\sigma$'s may differ for elastic scattering and mass conversion $\sigma_s(v)\not=\sigma_c(v)$. Here we show that the 2cDM model sets important constraints on DM properties, which can further be constrained by particle physics experiments. In order to study the DM dynamics in great detail, we restrict most of our simulations to the halo masses of less or about $10^{12}M_\odot$, i.e, at or below the MW-type mass. This sets the simulation box to be rather small ($L = 3 h^{-1}$Mpc). Thus, our simulations are not strictly cosmological but they allow for enough spatial and mass resolution with reasonable computing resources. In this paper, we explore the 2cDM parameter space and determine which $\sigma(v)$-models are in agreement with observations. Our study required the total of 57 simulations, so our approach is the most adequate for the task. We also compare 2cDM with other existing models, such as CDM and SIDM. 

In a series of papers, we will be numerically exploring how structure formation occurs within the novel $N$-component DM model. In this paper (Paper 1), we study the global statistics of DM halos and, particularly, their distribution by mass represented by their maximum circular velocity function. We demonstrate that the simplest two-component DM model is robustly resolving the substructure and too-big-to-fail problems. Subsequent papers will be devoted to detailed studies of dark matter halo structure and evolution across the mass range from dwarf galaxies to galaxy clusters. In particular, in Paper 2 (Todoroki $\&$ Medvedev, in prep) we investigate the structure and evolution of galactic halos within the multicomponent dark matter paradigm.

This paper is organized as follows. In Section \ref{sec:methods}, we present an overview of the theoretical formulation of the 2cDM model and describe the numerical techniques used. In Section \ref{sec:vmax_VSIDM} we show the model comparison and the parameter studies on the 2cDM model by looking at the maximum circular velocity function. The validity of each model is checked by comparing our results with observational data.  Section \ref{sec:vmax_evo} discusses the evolution of the velocity function. In Section \ref{sec:distribution} we discuss a novel (to our knowledge) small-scale problem of CDM related to the radial distribution of dwarfs in a parent halo. We demonstrate that 2cDM model is capable of resolving it too. In Section \ref{sec:spectrum} we discuss the matter power spectrum in 2cDM and its redshift evolution, e.g., as a proxy to the Ly-$\alpha$ forest. We summarize our conclusions in Section \ref{sec:CN}.


\section{Methods} 
\label{sec:methods}

We implemented the self-interacting 2cDM model in the TreePM/SPH code GADGET-3 \citep{springel2005, springel2008}. The numerical implementations closely follow those presented in \cite{medvedev2014} with upgrades and  optimizations. The model's detailed theoretical foundations are described in \cite{medvedev2010, medvedev2010b,medvedev2014theo}, and we present here the most important aspects of the models, for completeness. In the SPH simulations, dark matter, gas, and stars are all represented by simulation particles of different kind. Typically, gas is the only type of particle that interacts with each other both gravitationally and non-gravitationally through hydrodynamical forces. In the 2cDM model, however, we let dark matter particles interact non-gravitationally with each other, namely via both elastic scattering and mass conversion. In this paper, we present the results of $N$-body DM-only simulations. We postpone consideration of the dynamics of gas and stars and the role of baryonic physics to forthcoming papers.   

In the 2cDM model, two important physical processes are inherent for dark matter: elastic scatterings and mass eigenstate conversions. It is these physical processes that set 2cDM apart from the generic SIDM model \citep{spergel2000,yoshida2000,moore2000, colin2002,ahn2005,rocha2013}. Within the framework of 2cDM, the DM is postulated to consist of two mass eigenstates: {\it h} (`heavy') and {\it l} (`light'), which are represented in a code by standard simulation particles of different mass. They are allowed to scatter off each other ellastically or be converted from one another through inelastic scattering without creating/destroying additional particles. The previous study by \cite{medvedev2014} demonstrated that the difference in the masses of the two eigenstates is many orders smaller than the mean DM particle mass, namely $\Delta m /m \sim 10^{-7}-10^{-8}$. In this case $m_h\approx m_l\approx m$ and it is instructive to introduce a related parameter, the `kick velocity', $V_k=c\sqrt{2 \Delta m/m}$. Thus, $V_k\sim100$~km~s$^{-1}$ used in most simulations reported here (unless stated otherwise) corresponds to $\Delta m /m \sim 6\times10^{-8}$.

The initial composition, i.e., the ratio of the total numbers of each DM quantum species, $h$ and $l$, at the starting redshift, is taken to be 50:50, which is natural for fully decohered flavor-mixed particles. In general, the mass conversion between {\it h} and {\it l} can occur via multiple pair-wise processes ($hh \rightarrow ll, hh \rightarrow hl, hl \rightarrow ll$, etc) in which both energy and momentum are manifestly conserved. The process in which one (both) `heavy' is (are) converted into `light' is particularly interesting, because it effectively causes escape, called `quantum evaporation', of the DM secondaries from fairly small DM halos with the escape velocity below $V_k$. That is, in a simple $h \rightarrow l$ process, the resultant kinetic energy increases by $\Delta mc^2$. If, after conversion, the `light' DM particle's velocity exceeds the escape velocity of the associated gravitational potential, it breaks free and escapes, thus reducing the halo mass \citep{medvedev2010b}. The opposite process in which $l$ is converted into $h$ can also occur if it is kinematically allowed, i.e., if there is enough kinetic energy to create a heavy mass eigenstate from a lighter one. 

In the code, we use the Monte-Carlo technique for modeling DM self-interactions under the assumption of {\em rare binary collisions}. It is appropriate for a system of particles whose interaction probabilities are much less than unity within a dynamical time. The probabilities of the interaction processes that can occur during $\Delta t$ are computed as
\begin{equation}\label{eq:probability}
P_{ij \rightarrow i'j'} = (\rho_{j}/m_{j})\sigma_{ij\rightarrow i'j'}|{\rm \bf v_{\it j} - v_{\it i}}| \Delta t \ \Theta(E_{i'j'}),
\end{equation}
where $i$ and $j$ denote a `projectile' and a `target' particle respectively, $\rho_{j}/m_{j}$ is the number density of the target particle, $\sigma_{ij\rightarrow i'j'}$ is the velocity-dependent DM cross-section for the process $ij\rightarrow i'j'$, where the unprimed and primed indices denote initial and final states respectively,  ${\rm \bf v_{\it j} - v_{\it i}}$ is the initial relative velocity of the interacting pair, and $\Theta(E_{i'j'})$ is the Heaviside function that screens out kinematically forbidden processes with negative final kinetic energy $E_{i'j'} < 0$. Our implementation ensures that the probability of the vast majority of the interactions is kept below $\sim$0.001 for the rare binary collision approximation to be valid. However, we note that this approximation cannot be maintained for some extreme cases with large DM cross-section and strong velocity dependency of the DM interaction process. 

The implementation of Monte-Carlo interaction algorithm is as follows. At each time-step, (i) a pair of closest neighboring DM particles is identified; their types define the input channel, (ii) the probabilities of all four possible processes for a given input channel are calculated according to the $4\times4$ scattering matrix discussed by \cite{medvedev2014theo}, (iii) either one output channel of all possible ones or none of them is further chosen randomly in accordance with the computed probabilities, (iv) if an interaction occurs, the pair kinematics is computed in the pair's center-of-mass frame and the final kinetic energy and momenta magnitudes are computed from the energy-momentum conservation for the specific output channel at hand (e.g., the energy equal to $\Delta mc^2$ or twice as much is taken or given to the particles, depending on the process' type, that is whether one or both particles experience mass conversion, respectively); the final momenta directions are antiparallel and set randomly in the center of mass frame, and (v) once the pair had interacted, these particles are not allowed to interact again within the same time-step. 

The velocity-dependent cross-section is parametrized as 
\begin{equation} \label{eq:veldep}
 \sigma_{i\to f}(v) =     \left\{ \begin{array}{ll}
        \sigma_{0} (v/v_{0})^{a_{s}} & \mbox{for scattering,} \\ 
        \sigma_{0} (p_f/p_i) (v/v_{0})^{a_{c}} & \mbox{for conversion,} 
                \end{array}\right.
\end{equation}
where $p_i$ and $p_f$ are the initial and final momenta of the projectile particle, $v_{0}=100$~km s$^{-1}$ is the conventional velocity normalization and $\sigma_{0}$ is a common numerical coefficient. Although scattering and conversion cross-section values generally differ from each other, they are of the same magnitude for the maximal flavor mixing $\theta=\pi/4$ and similar but opposite flavor-interaction strengths $V_{\alpha\alpha}\simeq -V_{\beta\beta}, \ V_{\alpha\beta}\simeq V_{\beta\alpha}\simeq0$, see \cite{medvedev2014theo} for more details. The power-law indices $a_s$ and $a_c$ for elastic scattering and mass conversion, respectively, are treated independently. This allows us to choose a whole set of possible values for $a_{s}$ and $a_{c}$ to explore. Finally, the $(p_f/p_i)$ prefactor in the conversion cross-section is required to satisfy the quantum-mechanical detailed balance in the forward and reverse interaction probabilities:
\begin{equation}
\label{eq:sigmaprefactor}
\sigma_{i\to f}\, p_{i}^{2} = \sigma_{f\to i}\, p_{f}^{2} .
\end{equation}

We label our models by the pair of index values $(a_{s}, a_{c})$ and the values of $\sigma_0/m$ and $V_k$, where we assumed that $m_h\approx m_l=m$.
Among all the possibilities, ($a_{s}, a_{c}$) = $(-2,-2)$ is an interesting case based on the quantum mechanical argument that gives the maximum conversion probability as follows.   
Consider the initial and final states, $i$ and $f$, in a binary interaction. The cross section's dependency on the scattering and conversion can then be written as
\begin{equation}
   \left\{ \begin{array}{lcl}
       \sigma_{s}(v) = \sigma_{i \rightarrow i} \propto \ p_i^{-2}|1 - S_{ii}|^{2} \propto 1/v^{2} \\ 
       \sigma_{c}(v)  =  \sigma_{i \rightarrow f}  \propto \ p_i^{-2}|S_{if}|^{2} \  \propto  1/v^{2}
                \end{array}\right.
\end{equation}
where $S_{if}\equiv\langle f|\hat S|i\rangle$ is the scattering amplitude matrix.\footnote{Note that this index order differs from the conventional in quantum mechanics, $S_{fi}\equiv\langle f|\hat S|i\rangle$.} If initial and final states are the same, $S_{ii}$ describes elastic scattering, otherwise $S_{if}$ with $i\not=f$ describes inelastic interaction, i.e., mass conversion. The unitarity condition imposes that $\sum_\text{all final states} |S|^{2} = |S_{ii}|^{2} + \sum_f |S_{if}|^{2} = 1$, so that the conversion amplitudes $|S_{if}|=|S_{fi}|$ are at maximum if the pure scattering amplitude vanishes $S_{ii} = 0$,\footnote{This is true if both $i\to f$ and $f\to i$ are allowed; otherwise if $f\not\to i$ then $|S_{fi}|=0$, hence $\sum_f |S_{if}| = |S_{if}| = 1$, i.e., maximal as well.} which leads to maximizing $\sigma_{c}$. It also has a nice symmetry in respect of the velocity dependence and automatically satisfies the detailed balance condition, Eq. (\ref{eq:sigmaprefactor}). Thus, $\sigma_{s}(v)\simeq\sigma_{c}(v) \propto  1/v^{2}$. 

Another interesting case is ($a_{s}, a_{c}$) = $(0,0)$, because it is very natural since it represents the most common $s$-wave (`hard sphere') interaction of particles. 

The density of DM is computed by following the basic SPH formalism. For 2cDM, we distinguish heavy and light DM particles by simply tagging them as `$h$' and `$l$' since $\Delta m/m \sim 10^{-7} - 10^{-8}$, i.e., $m_{h} \approx m_{l}$, which means these two different DM species are numerically indistinguishable by mass in our simulations. The code estimates each DM species particle's density by first finding its neighbors by the oct-tree \citep{springel2005}. The number of neighbors for the target particle is set to 33 with the allowed deviation of $\pm2$. 
In the 2cDM model, we naturally have a mixture of heavy and light DM particles among the neighbors, and thus we compute the partial (or fractional) local DM mass densities for the heavy and light species separately as
\begin{equation}
\rho_{j, {\rm DM}} = \sum^{N}_{k=1} f_{k} m_{k} W(|{\rm \bf r_{\it k} - r_{\it j}}|, h_{j}), 
\end{equation}
where $j$ and $k$ denote the target and its $k$-th neighboring DM particles respectively,  $W$ is the SPH smoothing kernel function, {\bf r} is the position vector of the particle, and $h_{j}$ is the adaptive smoothing length of the target particle which encloses all of its neighbors in a spherical volume of space. 
The specific DM species among the neighbors is obtained by setting $f_{k}$ = 1 if the $k$-th neighbor is the same DM species as the target $j$, and otherwise 0.

The force resolution is set by the gravitational softening length, $\epsilon$, for each DM particle. It essentially `softens' the gravitational force in order to prevent the particle to experience unrealistic, too strong acceleration caused by neighboring particles in close proximity. Note also that the smoothing length $h_j$ is allowed to be reduced to 10$\%$ of the gravitational softening length of the particle (i.e., 0.1$\epsilon$) as the minimum value.

\begin{figure*}
  \centering
  \includegraphics[scale = 0.45]{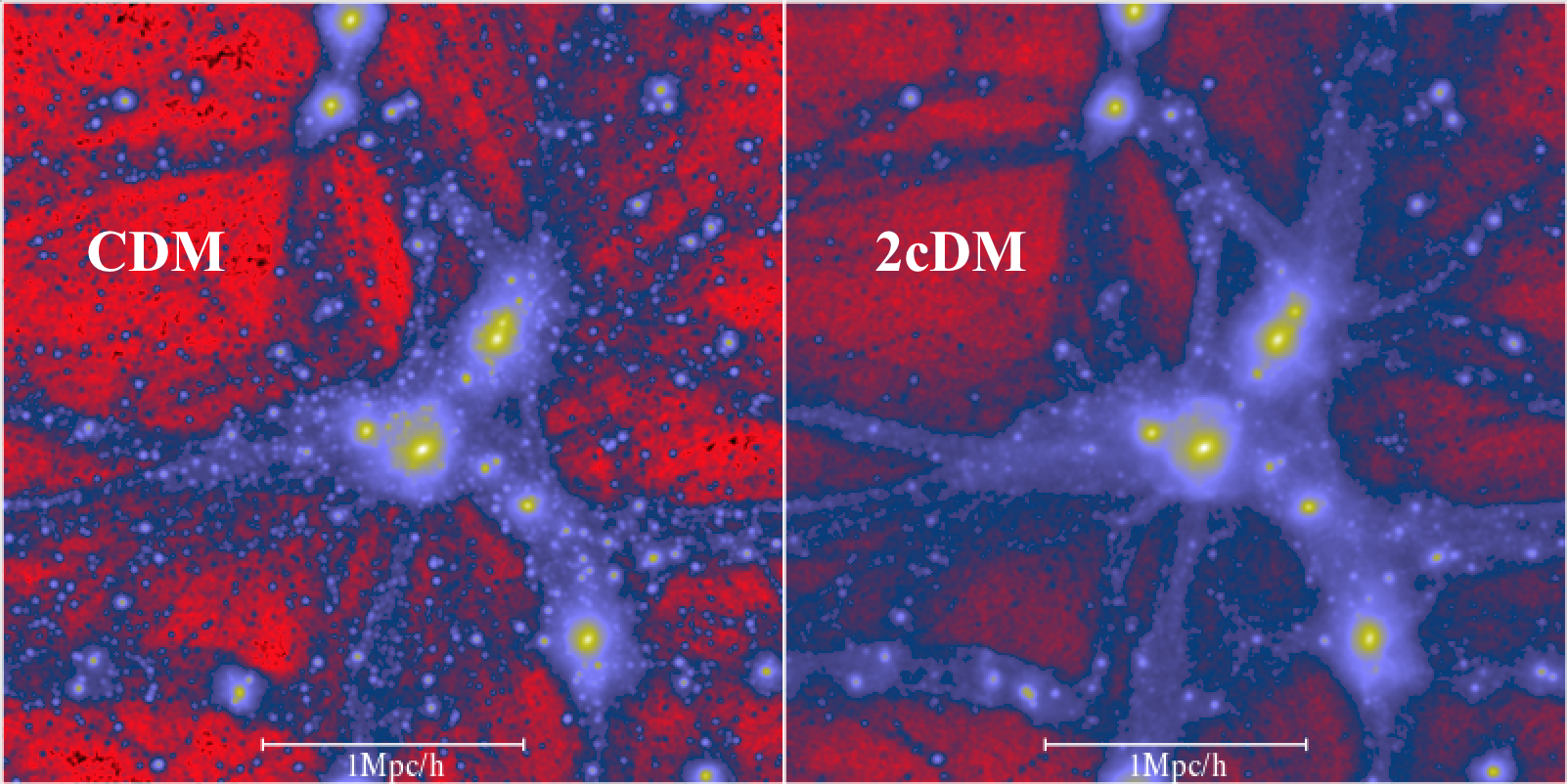}  
  \caption[]{\label{fig:viz} %
Dark matter density projection of the full box of $3h^{-1}$Mpc on a side, comparing CDM and a case from 2cDM at $z = 0$. The number of small satellite halos are reduced effectively by the 2cDM cosmology without altering the large scale distribution of DM.   
  }
\end{figure*}

The initial conditions were generated by the publicly available N-GenIC code with the same seed value to allow for reliable comparison across the models.
A periodic cube of side $3h^{-1}$ Mpc with the total number of 256$^{3}$ particles was used, where {\it h} is the normalized Hubble constant, $h=H_0/(100~\textrm{km~s}^{-1}\textrm{Mpc}^{-1})$, and the initial force resolution was set to 0.6 kpc. The largest and second-largest halos simulated in the box are on the order of $\textrm{a few}\times10^{11}M_{\odot}$, and they are roughly a factor of two smaller than the Milky Way and the Andromeda galaxies. Despite of the difference in the halo sizes, Figure~\ref{fig:viz} shows that our simulation box provides us with a good testing ground for the system that closely resembles the Local Group. We assume the set of cosmological parameters consistent with \cite{planck2015}: $\Omega_{m} = 0.31$,  $\Omega_{\Lambda} = 0.69$,  $\Omega_{b} = 0.048$, $\sigma_{8} = 0.83$, $ n_{s} = 0.97$, $h = 0.67$. The starting redshift was chosen to be $z_{i}=99$, and all simulations were carried to $z=0$.
For post-analysis, we use the Amiga Halo Finder (AHF) \citep{knollmann2009} to extract the halo properties and derive the profiles and mass functions. 

Our general strategy is to explore a very wide range of parameter sets for $\sigma(v)$ in Eq. (\ref{eq:veldep}). 
We then compare our results with observational data and constrains to determine the best fit parameters and possibly rule out some of the cases.   
For $\sigma(v)$ in our 2cDM model, we chose a range of parameters for (i) the cross-section per unit mass of $\sigma_{0}/m = 0.01, 0.1, 1$ and 10 cm$^{2}$g$^{-1}$ at $V_{k} = 100$~km~s$^{-1}$, and (ii) the $\sigma(v)$ power-law indices for the scattering and conversion, $(a_{s}, a_{c})$, each ranging over $0,-1$, and $-2$. They correspond to $s$-wave, $p$-wave and maximum-conversion (see above) interactions. We also checked $-4$ for $a_{s}$ only, which corresponds to the Rutherford-type long-range interaction cross-section. The power of $-3$ and anything smaller than $-4$ were not explored, due to the lack of physical motivation. 

Here we should note that there can be a potential numerical effect which can lead to the enhanced elastic scattering in the centers of dense halos. Here, a simulation particle can have a substantial interaction probability within a time shorter than the dynamical time in dense halo centers. In this regime, the role of elastic scattering is artificially enhanced, while that of mass conversions is diminished. To see why, let's consider a $hl\to ll$ conversion, for example. Once it happens in a small halo, the interacted particles have a large enough velocity to escape from the halo. That's what would have occured in nature. In simulations, however, a simulation particle represents a large ensemble of physical DM particles. Thus, it may happen that the interacted simulation particle experiences another conversion on its way out, $ll\to hl$, which reduces its velocity and prevents it from escape. This sequence of interaction processes $hl\to ll\to hl$ is dynamically similar to elastic scattering. Thus, the role of evaporation can be suppressed and the role of elastic scattering can be enhanced by this purely numerical effect. Such a process is expected to somewhat alter the inner parts of halos and potentially reduce the substructure suppression in large-scale cosmological simulations, where the mass resolution may be too coarse.


\section{Halo maximum circular velocity function}
\label{sec:vmax_VSIDM}

The halo maximum circular velocity function, or simply the velocity function, represents the cumulative distribution of the number of halos vs. their maximum circular velocities $V_{\rm max}$. 
The velocity function readily illustrates the two of the three problems, namely the SS and TBTF problems. The earlier work by \cite{klypin1999} showed that there is a large discrepancy between the observed circular velocity function of the Local Group and the one predicted from cosmological \lcdm numerical simulations. They used the observed data from \cite{mateo1998} and compared the velocity function with their $N$-body numerical simulations. The discrepancy is twofold: (i) \lcdm simulations over-predict the number of small halos in Local Group-type environments (the SS problem), and (ii) halos with $V_{\rm max}$ ranging from roughly 30 to 50 km s$^{-1}$ are absent in the observed Local Group, although they are massive enough to have standard star formation and, thus, must be observed, if present (the TBTF problem). The possible solutions for the discrepancy in their results were left inconclusive as to whether it is due to the failure of the CDM, or simply our ignorance of the detailed roles of gas and stellar dynamics, or possibly the undiscovered ultra-faint dwarfs that have not been accounted for by observations.

More recent observations with improved resolution detected ultra-faint dwarfs in the Local Group, making the lower end of the velocity function steeper \citep{simon2007}. These newly discovered dwarf satellites tend to reduce the discrepancy at the very low mass end ($V_{\rm max} \lesssim 10$ km s$^{-1}$), but the discrepancy seen in the `intermediate' mass scales remains strong. This non-negligible difference at the intermediate mass scales that cannot be explained by discovering more ultra-faint dwarfs is at heart of the TBTF problem. It appears that neither improvements of observations nor building larger, more complete sets of observational data could be the immediate solution to it.

From a theoretical aspect, considering the success of the CDM paradigm in reproducing the large-scale structure of the Universe, one might naturally consider baryonic feedback, or more specifically stellar feedback, to be one of the possible solutions to the above problems. It is, however, questionable whether including stellar feedback to the CDM framework can reasonably alter the halo populations in such a way that both the SS and TBTF problems are consistently resolved. For example, \cite{garrison-kimmel2013} found that even a strong supernova feedback with rather unrealistically high energy output cannot satisfactorily solve the TBTF problem, but see more recent simulations by \citet{wetzel2016}. 
Regardless of this, any baryonic feedback implemented in simulations are subject to a numerical resolution limit and the results can be highly model-dependent. Aside from relying on baryonic physics, \cite{garrison-kimmel2014} found that a particular choice of $\sigma_{8}$, a cosmological parameter a value of which different surveys such as WMAP-7 and Planck results suggest slightly different values, does not seem to resolve the problem either.

In this section, we will demonstrate that the 2cDM model does not necessarily require the baryonic feedback to bring the velocity function in good agreement with observations. To elucidate the robustness of 2cDM, we also compare the results with other models, such as CDM and SIDM, which cannot resolve SS and TBTF. We then explore the 2cDM parameter space, namely different sets of ($a_{s}, a_{c}$) and $\sigma_{0}/m$, on the velocity function to see which model shows agreement with observational expectations. 

\begin{figure*}
\centering
\includegraphics[scale = 0.55]{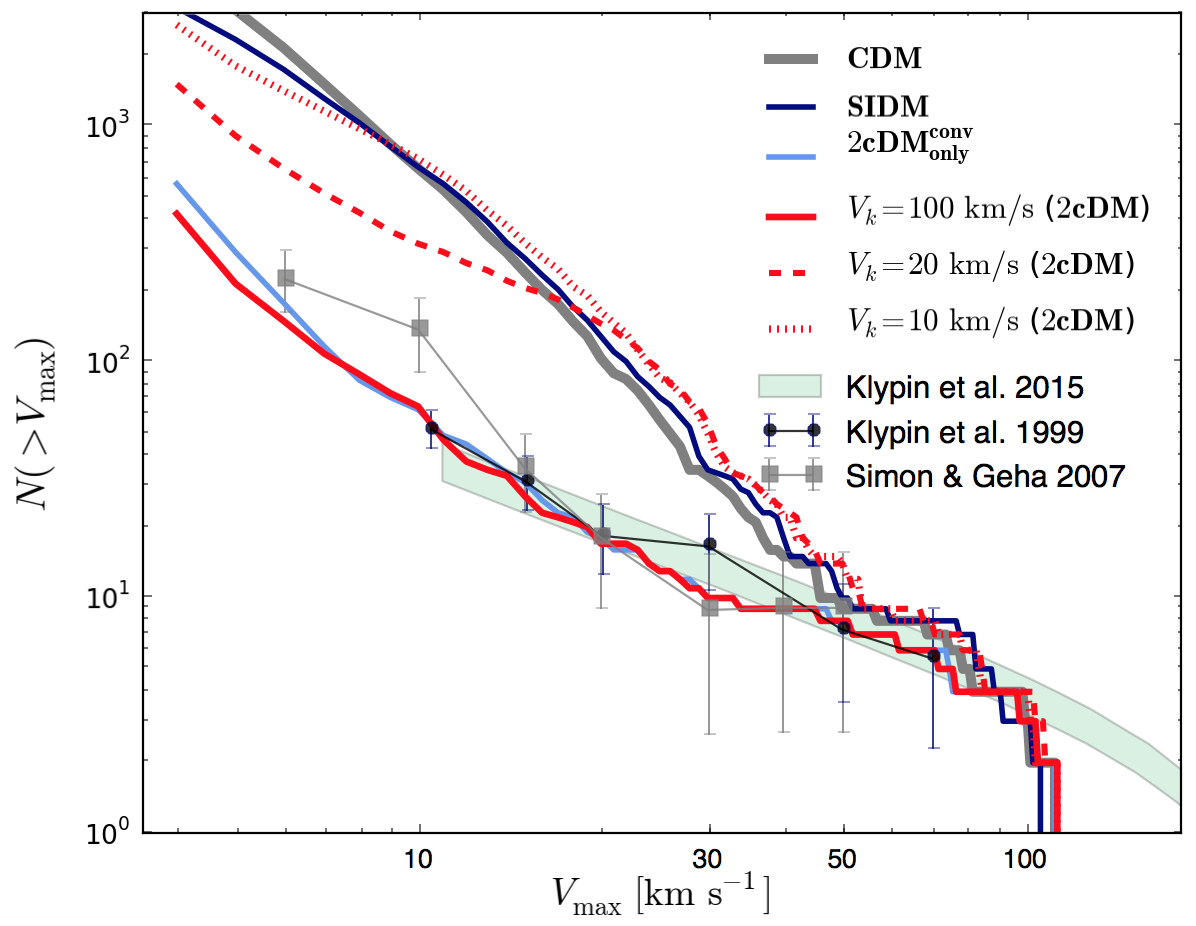}  
\caption[]{\label{fig:vmax_VSIDM} 
Maximum circular velocity functions (velocity functions) for the classical CDM, SIDM and various 2cDM models with $\sigma_{0}/m = 1$ and the $(a_s,a_c)=(-2,-2)$. Comparison of CDM (grey) and SIDM (dark blue) demonstrates that SIDM is unable to resolve the substructure and too-big-to-fail problems. Red curves (solid, dashed, and dotted) show that the break in the velocity function is set by the parameter $V_k$, which is assumed to be 100~km~s$^{-1}$ for the fiducial value. The comparison of the full 2cDM (solid red) and 2cDM without elastic scatterings (light blue) confirms that elastic interactions play no role in shaping the velocity function. The black and grey points with error-bars show the Local Group observational data from \citet{klypin1999} and \citet{simon2007}, and the data in the green strip is taken from \citet{klypin2015} for the abundance of field galaxies, with the number of halos properly normalized as in \citet{medvedev2014}. 
}
\end{figure*}

Figure~\ref{fig:vmax_VSIDM} highlights the distinction between the models. Here we present the velocity function for CDM, SIDM and 2cDM models compared with \cite{klypin1999,klypin2015} and \cite{simon2007} data. Obviously, CDM (grey curve) is largely inconsistent with observations. SIDM, which was proposed to resolve the CC problem, shows no suppression of the number of small halos either, and thus it fails to resolve the SS and TBTF problems.\footnote{The only difference seen between SIDM and CDM is the minor reduction of satellite halos in the lower-mass end, which we attribute to a numerical effect on physics grounds. }
This is because elastic collisions {\em redistribute} DM particle velocities but neither change the total kinetic (`thermal') energy of particles in the halo, which thus remains in a virial equilibrium, nor they fully thermalize the halo to make thermal evaporation appreciable. 
Hence the virial mass of a DM halo is generally unaffected in SIDM for reasonable values of the cross-section parameter studied in this work ($\sigma_{0}/m = 0.01$ to 1 cm$^2$ g$^{-1}$).

On the other hand, the effect of mass conversions on the velocity function is very prominent. One of the 2cDM model parameters, $V_k$ --- the ``kick velocity" due to DM conversions --- sets the position of the break in the velocity function, because only the halos with their escape velocities smaller than $V_k$ are subject to evaporation. Figure \ref{fig:vmax_VSIDM} illustrates this with the velocity functions for $V_{k} = 10$, 20 and 100 km s$^{-1}$. These results confirm previous studies. Note that the exponential cutoff at $V_{\rm max}\sim100$~km~s$^{-1}$ is artificial due to the smallness of the simulation box. It is of no concern to us since our goal here is to study the low mass (or low $V_{\rm max}$) end of the cumulative halo counts, and because the CDM and 2cDM velocity functions match at $V_{\rm max}\sim100$~km~s$^{-1}$. In order to illustrate that it is the mass conversions rather than elastic collisions that suppress the velocity function, we present the velocity function for the simulations without elastic collisions (physically, this is impossible due to the unitarity condition in scattering), which is labeled as 2cDM$^{\rm conv}_{\rm only}$. One can see that this conversion-only case shows a strong suppression of the overabundant satellite halos nearly identical to the full 2cDM model and is grossly different from SIDM. Obviously, the 2cDM halo velocity function closely matches the observed one, and therefore the 2cDM model successfully resolves both the SS and TBTF problems. 

In our study presented below, we compare our results with observations on the Local Group \citep{klypin1999, simon2007} and the Local Volume \citep{klypin2015}. As is explained in Section \ref{sec:methods}, the simulation box we chose ($L=3h^{-1}$Mpc) well represents the Local Group-type of environment, which is statistically analogous and directly comparable to the observed Local Group data. Meanwhile, our simulations are primarily aimed at studying the low-mass (low-$V_{\rm max}$) end. Therefore, even though the Local Volume data compiled by \cite{klypin2015} has a larger statistical sample than our simulations, we can still compare the slope of the low-mass end. In fact,  the 2cDM model and the observations show a remarkably good match at the low-mass end, whereas the CDM and SIDM both fail badly.

\begin{figure*}
\centering
\includegraphics[scale = 0.6]{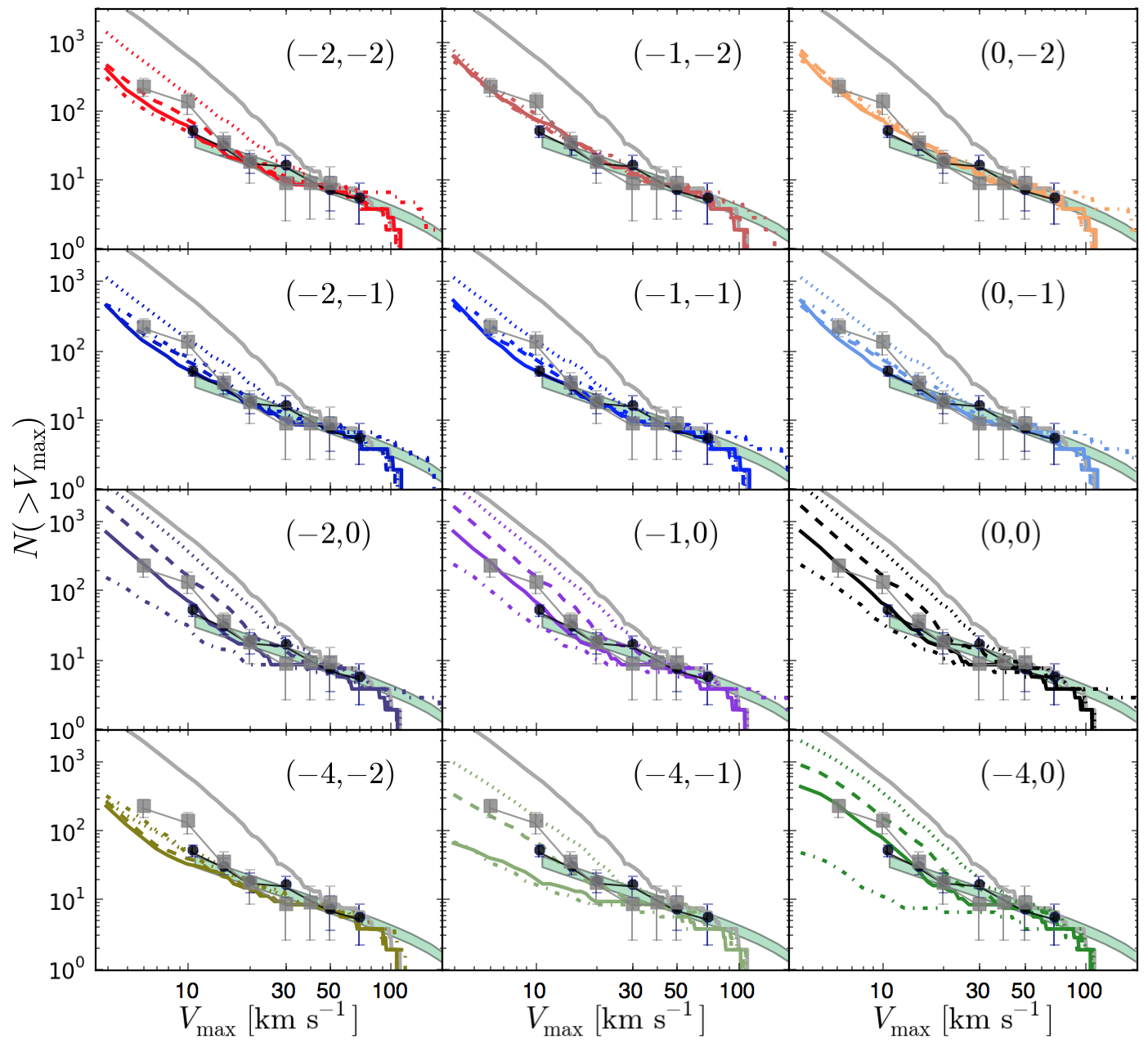}  
\caption[]{\label{fig:vmax_all} %
Maximum circular velocity functions for 2cDM with various $\sigma(v)$-models. The thick solid gray curve is CDM. The dotted, dashed, solid, and dash-dot curves correspond to $\sigma_{0}/m = 0.01,0.1,1$ and 10, respectively. The legend for the observational data points are the same as in Figure~\ref{fig:vmax_VSIDM}.
}
\end{figure*}

We now explore the parameter space available for the 2cDM model with the velocity function. In Figure~\ref{fig:vmax_all} we compare different sets of 2cDM parameters (which we shortly call `2cDM models') with CDM and observations. Overall, we find that 2cDM copes with both the SS and TBTF problems remarkably well for the most of the parameter sets. We point out and elucidate some of the features that are unique to 2cDM in the following.

We first remind the reader that the case ($-2,-2$) does {\em not} have a ($p_f/p_i$) prefactor (or the `$\sigma$-prefactor', which is $\propto1/v$ for $v\ll V_k$) to the  $\sigma(v)$ in Eq. (\ref{eq:veldep}), unlike all the other cases. 
The key features of how $a_{s}$ and $a_{c}$ affect the velocity function can easily be seen by comparing the panels row by row. Each row, except for the bottom row, represents the role of the scattering power-law index $a_{s}$ on shaping the velocity function while holding the mass conversion power-law index $a_{c}$ unchanged. The first, second and third rows show that the scattering index $a_{s}$ has little effect on the velocity function (i.e., left, middle and right panels in each row are nearly identical), except for ($-2,-2$) which has no $\sigma$-prefactor. 
Furthermore, when comparing the first three rows, it is interesting to see that the mass conversion power-law index $a_{c}$ has an effect of reducing the velocity function's dependence on the magnitude of $\sigma_{0}/m$.
In other words, as one goes from $a_c=0$ (third row) to $a_c=-2$ (first row), the differences seen among the cases with different $\sigma_{0}/m$ values diminish and converge. 
To summarize, the controlling agent for the velocity function is the mass conversion power-law index $a_{c}$ rather than that of scattering $a_{s}$. This conclusion can also be established by comparing the panels column by column, for each column has the velocity functions behaving identically regardless of the scattering power-law index $a_{s}$.  

The cases with $a_{s} = -4$, which correspond to Coulomb-like $\sigma_s\propto 1/v^4$ scattering, have also been explored and they are presented in the bottom (fourth) row. We stress that caution must be taken when comparing $a_{s} = -4$ with the rest of the other cases, especially for larger $\sigma_{0}/m$ values. As we noted earlier, the 2cDM numerical modeling uses the rare binary collision approximation and assumes DM to be weakly interacting. The direct consequence of $a_{s} = -4$ is that it increases the number of DM interactions significantly in low-mass, low-velocity halos and puts them in a strongly interacting fluid-like regime. The number of interactions also depends on $\sigma_{0}/m$, which itself causes the similar but weaker effect compared to $a_{s}$. Consequently, a combination of $a_{s} = -4$ and larger values of $\sigma_{0}/m = 1$ or 10 amplifies the interaction probabilities dramatically and cannot be reliably used for comparison with the other cases.

In addition to the strong influence of the conversion power-law index $a_{c}$ described above, there are general features that are commonly seen in most of the 2cDM cases as follows:

(i) The magnitude of the DM cross-section $\sigma_{0}/m$ determines the degree of suppression of the number of satellites, hence offering a solution to the SS and TBTF problems. In most cases, the larger the cross-section, the stronger the velocity function suppression. Interestingly, for  $a_{c}$ being $-1$ and $-2$, the difference among various cross-sections, $\sigma_0/m$, is rather minimal, indicating strongly nonlinear effects of DM interactions, gravitational collapse, accretion and hierarchical merging.

(ii) Many 2cDM $\sigma(v)$-cases robustly reduce the halo counts with $V_{\rm max} \lesssim 50$~km~s$^{-1}$ (corresponding to $M_{\rm vir} \lesssim 10^{9.5}M_{\odot}$) to be in agreement with observations. One of the key parameters, $V_k$, which sets the position of the break, is also set by observations \citep{medvedev2014} to be $V_k \sim 50-100$~km~s$^{-1}$. Such a value of $V_{k}$ corresponds roughly to $\Delta m/m \sim 10^{-7} - 10^{-8}$.

(iii) Our results set constrains on the magnitude and velocity-dependence of the DM self-interaction cross-section. Namely, $\sigma_{0}/m = 0.1$ and 1 appear to be consistent with observations for most of the $\sigma(v)$-models. The smaller cross-section of $\sigma_{0}/m = 0.01$ are still in agreement with observations for the two cases of ($-1,-2$) and ($0,-2$). The larger cross-section of $\sigma_{0}/m = 10$ is ruled out because it tends to over-suppress dwarf halos and produces an even larger number of large-mass halos than the CDM counterpart. We should note that baryonic feedback effects may play an additional important role in reducing the overall magnitude of the velocity function as well. Therefore, we cannot rule out the small cross-sections ($\sigma_{0}/m = 0.01$) unless we test 2cDM with baryonic feedback. This will be explored in forthcoming studies. 

Although the reported results are robust across the parameter domain, we should take the results below $v\sim 10$~km~s$^{-1}$ with caution as they may be affected to some extent by numerical resolution effects. Recently, \cite{power2016} proved that both CDM and WDM simulations are not immune from the formation of spurious small-scale halos. They argued that the problem is intrinsic to the codes themselves and the discreteness-driven relaxation is hard to eradicate even by an appropriate choice of the gravitational softening length. In a subsequent study aimed at overcoming this issue, \cite{hobbs2016} developed an adaptive softening scheme that minimizes anisotropic distribution of particles/cells, leading to a reduction in the spurious sub-structures. Such a scheme is not present in GADGET code used here. Based on this argument, we expect that our results can slightly over-predict the halo counts at the very lower-mass end. 

Finally, to quantify our findings, we report the cumulative number of halos, i.e., the value of the velocity function, evaluated at $V_{\rm max} = 15$ km s$^{-1}$ --- well inside the observed data from \citet{simon2007}, hence avoiding possible observational biases. Table~\ref{table:V15} summarizes and compares our simulation results with observations. The $\sigma(v)$-models that are consistent with observations within 1$\sigma$ error bars are shown in bold. One sees that the majority of the models fall within the observed uncertainty range.

\begin{table}
\centering
\tabcolsep=0.08cm
\begin{tabular}{ccc|ccc|ccc|ccc}
\hline\hline
Model & $\sigma_{0}/m$ & $N$ & Model & $\sigma_{0}/m$ & $N$ & Model & $\sigma_{0}/m$ & $N$ \\
\hline

$(-2,-2)$ & 0.01 & 79  & $(-1,-2)$ & 0.01 & {\bf 42} & $(0,-2)$ & 0.01 & {\bf 44}  \\
           & 0.1   & {\bf 35}  &	   & 0.1 & {\bf 39}    &    & 0.1 &  {\bf 37} \\
 	   & 1      & {\bf 27}  & 	  & 1     & {\bf 38}     &   & 1   & {\bf 38}  \\
           & 10     & {\bf 29} & 	  & 10   & {\bf 44}      &   & 10  &  {\bf 39}   \\[1ex]
$(-2,-1)$ & 0.01 & 73  & $(-1,-1)$ & 0.01 & 74 & $(0,-1)$ & 0.01 & 74  \\
           & 0.1   & {\bf 34}  &	   & 0.1 & {\bf 34}    &    & 0.1 &  {\bf 32} \\
 	   & 1      & {\bf 34}  & 	  & 1     & {\bf 34}     &   & 1   & {\bf 35}  \\
           & 10     & {\bf 42} & 	  & 10   & {\bf 44}      &   & 10  &  {\bf 45}   \\[1ex]
$(-2,0)$ & 0.01 & 141  & $(-1,0)$ & 0.01 & 145 & $(0,0)$ & 0.01 & 141  \\
           & 0.1   & 66  &	   & 0.1 & 70    &    & 0.1 &  69 \\
 	   & 1      & {\bf 27}  & 	  & 1     & {\bf 25}     &   & 1   & {\bf 28}  \\
           & 10     & 15 & 	  & 10   & 16      &   & 10  &  16   \\[1ex]         
$(-4,-2)$ & 0.01 & {\bf 34}  & $(-4,-1)$ & 0.01 & 79 & $(-4,0)$ & 0.01 & 141  \\
           & 0.1   & {\bf 25}  &	   & 0.1 & {\bf 29}    &    & 0.1 &  73 \\
 	   & 1      & {\bf 23}  & 	  & 1     & 15     &   & 1   & {\bf 31}  \\
           & 10     & {\bf 27} & 	  & 10   & 12      &   & 10  &  8   \\[1ex]                   
\hline\hline

\end{tabular}
\caption[]{ Cumulative number of halos with $N(>V_{\rm max} = 15\textrm{ km s}^{-1})$ for all the cases. The normalized $N$ at $V_{\rm max} = 15$ km s$^{-1}$ from observation taken from \citet{simon2007} gives 35.6$\pm$13.1, which sets the observationally acceptable range of [$N_{\rm min}$, $N_{\rm max}$] = [22.6, 48.7]. The bold-faced are the ones that fall in this range. }
\label{table:V15}
\end{table}


\section{Evolution of halo maximum circular velocity function} 
\label{sec:vmax_evo}

\begin{figure*}
\centering
\includegraphics[scale = 0.6]{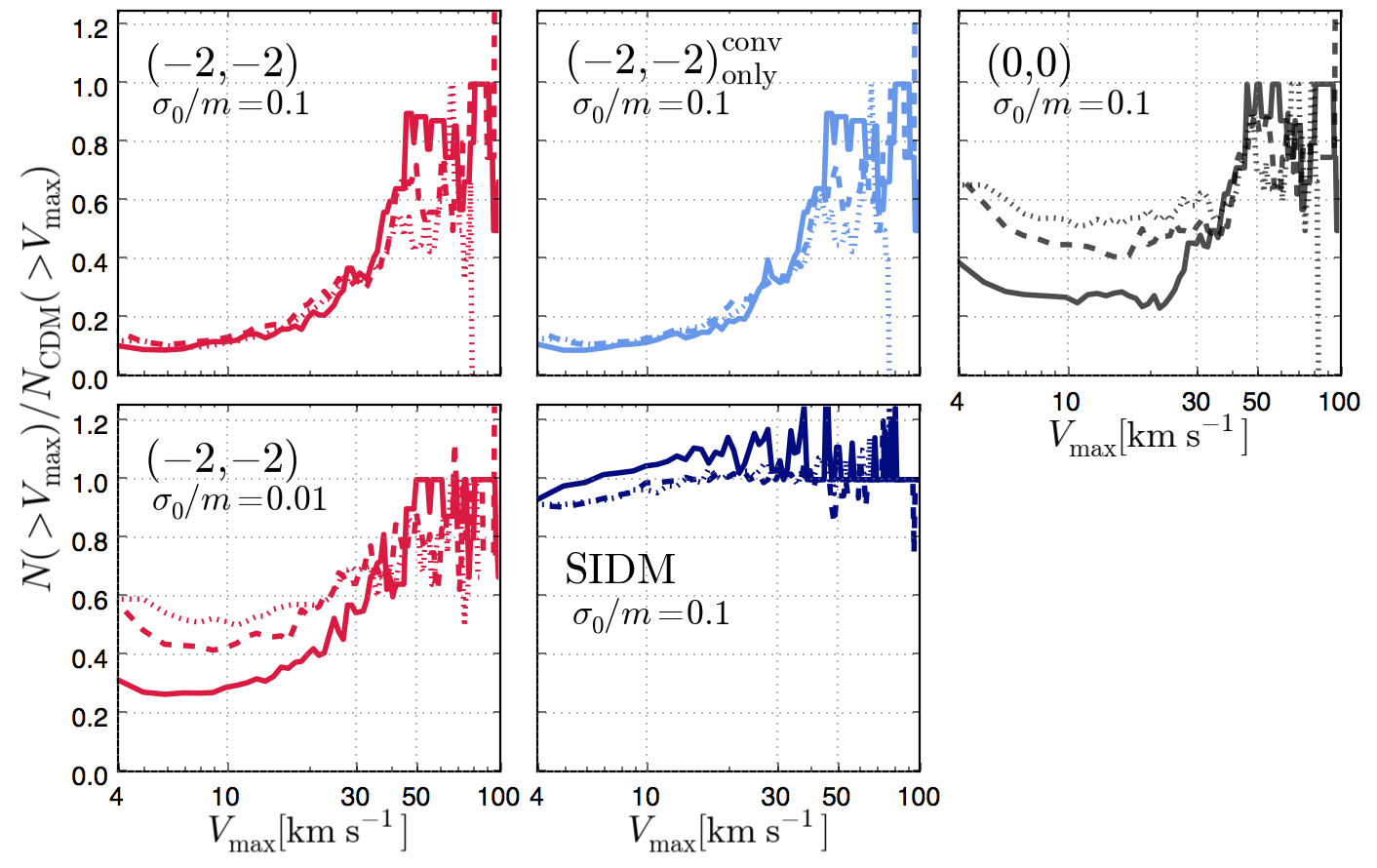}  
\caption[]{\label{fig:vmax_evo} %
Time evolution of the velocity function, normalized to the CDM counterpart. 
Solid, dashed, and dotted curves are for $z = 0, 2, $ and 4, respectively. 
}
\end{figure*}

In this section, we study the evolution of the halo velocity function with redshift. 
From a numerical point of view, the halo mass, i.e., the virial halo mass $M_{\rm vir}$, is more reliable than the maximum circular velocity of halo, $V_{\rm max}$, which is computed based on the mass distribution within the halo. On the other hand, $M_{\rm vir}$ depends on the definition of the halo virial radius, 3which is somewhat arbitrary. Furthermore, in the observational data the estimated $V_{\rm max}$ is more trustworthy than the total mass of the halo since estimating the halo mass requires additional assumptions on the density profile and shape of the halo. 
In principle, both the velocity and mass functions deliver the same type of information regarding the halo abundance with respect to the size of the halos. 
Here we focus on the velocity function to go parallel with observations. 
For the purpose of elucidating the correspondence of $V_{\rm max}$ and $M_{\rm vir}$, however, we also discuss the mass function when necessary. 

\subsection{Redshift dependence} 
\label{sec:vf_z}

In Figure~\ref{fig:vmax_evo} we compare the evolution of the 2cDM and CDM velocity functions as a function of redshift: $z=4$ (dotted), 2 (dashed), and 0 (solid). To elucidate the difference from the CDM model, the velocity function of each 2cDM case was normalized to that of the CDM counterpart. 
We first compare $(-2,-2)$ with $\sigma_{0}/m = 0.01$ and 0.1 in the left-most column.
We find that with $\sigma_{0}/m = 0.1$, the cumulative number of subhalos with $V_{\rm max} \lesssim 20$ km s$^{-1}$ can be reduced to $\sim10\%$ of CDM since the redshift of as early as $z=4$. For a smaller $\sigma_{0}/m = 0.01$, the suppression is weaker especially at higher $z$, and the cumulative number of halos is about $\lesssim 60\%$ of CDM at $z=4$ and $\sim 30\%$ at $z=0$. 
In both cases, the steep decline of halo counts, i.e., the deviation from the CDM halo counts, starts to occur at $V_{\rm max} \sim 50$ km s$^{-1}$, which corresponds to roughly M$_{\rm vir} \lesssim 10^{10}$M$_{\odot}$ in the mass function. The mass functions exhibit similar trends.

To shed light on how elastic scattering and mass conversion play a role in suppressing the substructures at higher $z$, we also show the mass conversion-only case, namely $(-2,-2)^{\rm conv}_{\rm only}$, and the scattering-only case, or simply SIDM, which are presented in the middle column. 
The former is nearly identical to the full $(-2,-2)$ case with both scattering and mass conversion enabled, and this is seen even at higher redshift of at least $z > 4$. On the contrary, the latter (i,e., SIDM) closely resembles CDM at $0 \leq z \leq 4$, indicating that elastic scattering itself does not make any significant difference in the velocity function at higher $z$ compared to the CDM model. (Note that SIDM here has the cross-section's velocity dependence of $v^{a_{s}}$ with $a_{s} = -2$. ) In turn, it means that the mass conversion process is the primary reason for the deviation of the velocity function from the predictions of the CDM model. The deviation of 2cDM from CDM starts to occur noticeably around $V_{\rm max} \lesssim 50$ km s$^{-1}$, primarily due to the velocity kick parameter $V_{k}$ that is set to 100 km s$^{-1}$ in all the cases presented in Figure~\ref{fig:vmax_evo}. Since this parameter is specific to the mass-conversion process, SIDM does not show any  drastic deviation from the CDM model at that $V_{\rm max}$ value. 

Thus far, we discussed the 2cDM model with $(-2,-2)$. Another interesting 2cDM case is $(0,0)$ with $\sigma_{0}/m=0.1$ in the top right panel. Among all, the $(0,0)$ case is particularly interesting since its cross-section has no velocity dependence, which is the most conventional model of interactions. At a high redshift, where the typical (virial) velocities of DM particles are smaller, the cross-section's velocity dependence makes a significant difference in the self-interaction rate. For example, the $(-2,-2)$ cases indicated that the smaller the velocities, the larger the interaction probabilities due to its negative power-law of $-2$. Hence, $(0,0)$ has a weaker suppression of substructures at higher $z$ compared to $(-2,-2)$. This is clearly manifested in Figure~\ref{fig:vmax_evo} where the strength of the substructure suppression for $(0,0)$ with $\sigma_{0}/m = 0.1$ is found to be roughly a factor of 5 to 6 weaker than that of the $(-2,-2)$ counterpart at $z \gtrsim 2$.
Interestingly, the ($0,0$) case suppresses the abundance of low-mass halos with $V_{\rm max} \lesssim 30$ km s$^{-1}$, or correspondingly $10^{7} \lesssim {M_{\rm vir}/M_{\odot}} \lesssim 10^{9}$, to roughly $60\%$ of CDM over $0 \leq z \leq 4$ consistently. 
This is in contrast to the steep velocity dependence of ($-2,-2$) cases with larger cross-sections of $\sigma_{0}/m \gtrsim 0.1$ in which the suppression becomes stronger for smaller halos due to their smaller velocity dispersions. 

It is also indicated that in order for a cosmological model to match the observed substructure abundance at present time and solve the missing satellite, or SS, and TBTF problems, a significant suppression of substructures, especially in terms of halo mass, may need to commence as early as $z \gtrsim 4$. We argue that the presence of baryons and stellar feedback effects at earlier times will unlikely do the work since the energy budget at such higher $z$ would not be sufficient to trigger a strong suppression of substructures in a systematic manner throughout the Universe. Meanwhile, the UV background has been suggested and shown to have positive effects on substructure suppressions in numerical works \citep{efstathiou1992, summerville2002, hambrick2011}, although there are other studies that show the effects are not significant \citep{quinn1996}. It is therefore possible that our $(-2,-2)$ model may have a too strong suppression when combined with the UV background at higher $z$. Needless to say, this is still dependent on the cross-section value and the power-law indexes of the cross-section's velocity dependence power-law. A more thorough study with baryonic physics should be able to provide insights into this.

\subsection{Scaling relations} 
\label{sec:mf_power_fit}

To quantify the degree of substructure suppression over $0 \leq z \leq 4$, we fit the lower-$V_{\rm max}$ and -$M_{\rm vir}$ ends of velocity and mass functions for the models presented in this section with a power-law. We chose the fitting ranges to be $4 \leq V_{\rm max}  \leq 30$~km~s$^{-1}$ and $10^{7} \leq M_{\rm vir} \leq 10^{10}M_{\odot}$, where the functions show a clear power-law with minimal effects of numerical uncertainty.
The power-law indices are defined as $b_{V_{\rm max}} \equiv d\ln N(>V_{\rm max}) / d\ln V_{\rm max}$ and $b_{M_{\rm vir}} \equiv d\ln N(>M) / d\ln M$. 
We briefly summarize the key results as follows.

In general, nearly all cases consistently show $b$ becoming shallower from $z=4$ to 0 for both velocity and mass functions although the changes can be considered minor. 
The exception is ($-2,-2$) with $\sigma_{0}/m = 0.1$, in which $b$ remains nearly constant over $z \lesssim 4$: $b_{V_{\rm max}} \sim -2.0$ and $b_{M_{\rm vir}} \sim -0.7$. With $\sigma_{0}/m = 0.01$, $(-2,-2)$ shows $b_{V_{\rm max}} \sim -2.4 \rightarrow -2.2$ and $b_{M_{\rm vir}} \sim -0.9 \rightarrow -0.8$.
Hence, a larger cross-section yields a shallower $b$ within the same model.
This is not surprising because smaller halos are preferentially suppressed when the number of DM interactions increases at later times.
To summarize, the $(-2,-2)$ 2cDM models exhibit weak dependence of the power-law indices with redshift over $0\lesssim z \lesssim 4$ and obey the scaling relations:
\begin{equation}
\begin{split}
N_{(\sigma_{0}/m) = 0.01} \propto & V_{\rm max}^{-2.3} \propto M_{\rm vir}^{-0.8}, \\
N_{(\sigma_{0}/m) = 0.1} \propto & V_{\rm max}^{-2.0} \propto M_{\rm vir}^{-0.7}
\end{split}
\end{equation}
within the fitting range of $4 \leq V_{\rm max}  \leq 30$~km~s$^{-1}$.
For reference, CDM shows $b_{V_{\rm max}} \sim -2.6$ with minor variations and $b_{M_{\rm vir}} \sim -1.0 \rightarrow -0.9$ from $z = 4$ to 0. That is, for a good estimate $N_{\rm CDM} \sim V_{\rm max}^{-2.6} \sim M_{\rm vir}^{-0.9}$ at $0\lesssim z \lesssim 4$.
We note that even though we performed the fit over limited ranges due to the small simulation box size, the amount of substructure is large enough to draw statistically significant conclusions on the values of $b$.

\section{Radial halo distribution profile}
\label{sec:distribution}

\begin{figure*}
\centering
\includegraphics[scale = 0.6]{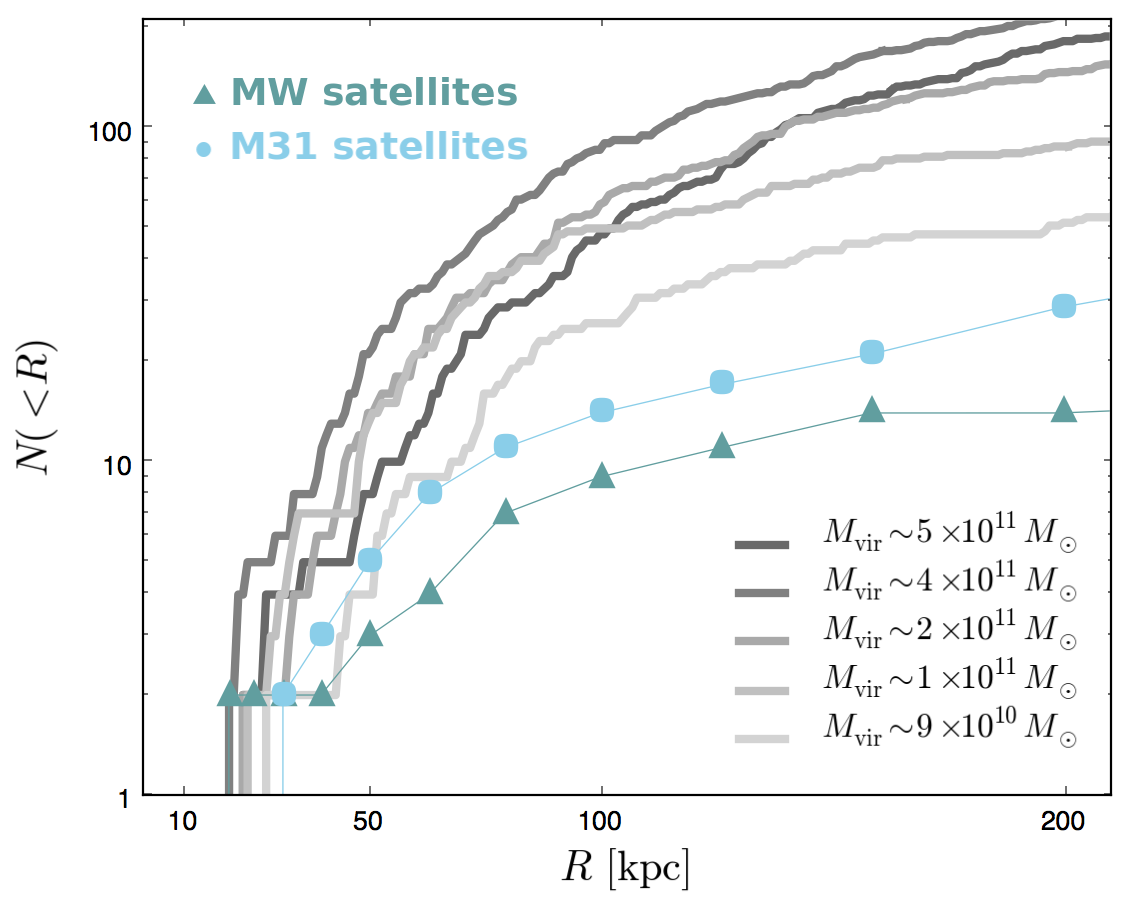}  
\caption[]{\label{fig:RHDP_CDM} %
Cumulative number of subhalos as a function of the host halo-centric radius: CDM vs. observations. Only subhalos with $M \geq 10^{7}M_{\odot}$ and $V_{\rm max} \geq 4$ km s$^{-1}$ are shown. The observed satellites satisfy the same mass cutoff at $M \gtrsim 10^{7}M_{\odot}$.
}
\end{figure*}

\begin{figure}
\centering
\includegraphics[scale = 0.43]{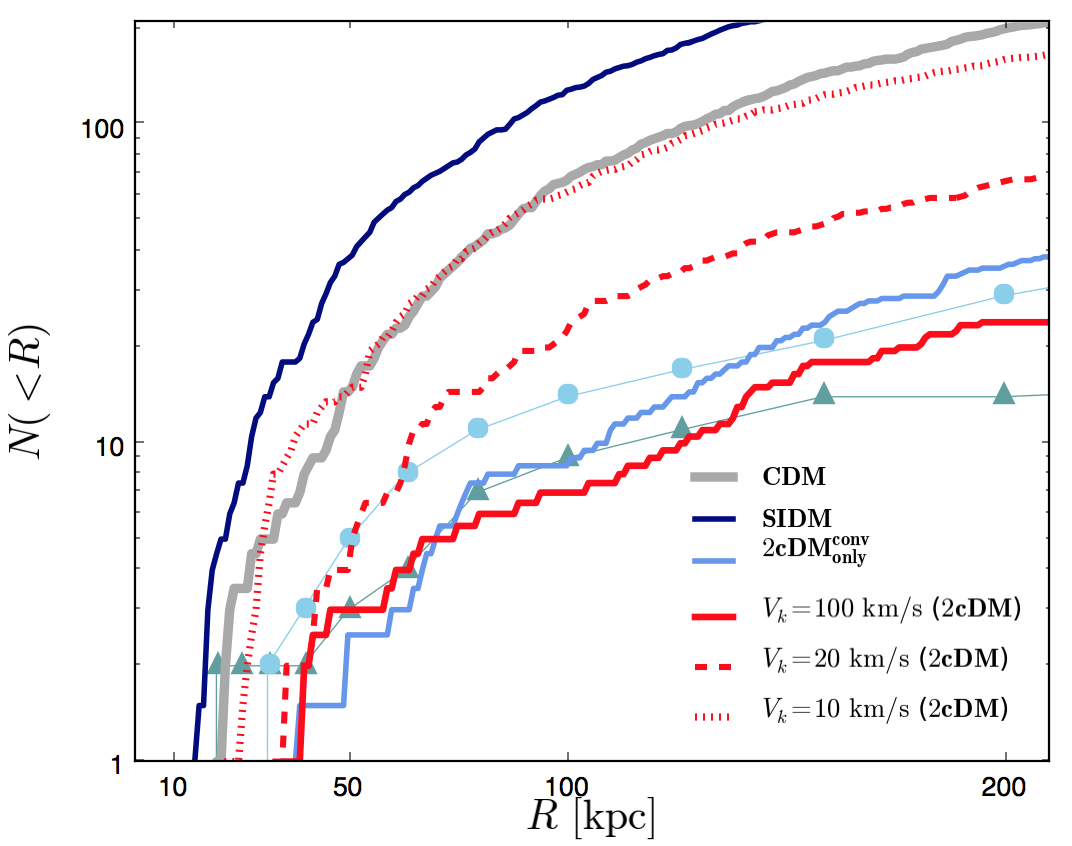}  
\caption[]{\label{fig:RHDP_VSIDM-sig1} %
Illustrative example of the cumulative number of subhalos as a function of the host halo-centric radius: CDM, SIDM and 2cDM vs. observations. The models are the same as in Figure \ref{fig:vmax_VSIDM}. All the cases from simulations are the averaged profiles of the two largest halos with $M_{\rm vir} \sim 5 \times 10^{11}$ and $4 \times 10^{11}M_{\odot}$. The same halo selection criteria have been applied as in Figure~\ref{fig:RHDP_CDM}. }
\end{figure}

\begin{figure*}
\centering
\includegraphics[scale = 0.6]{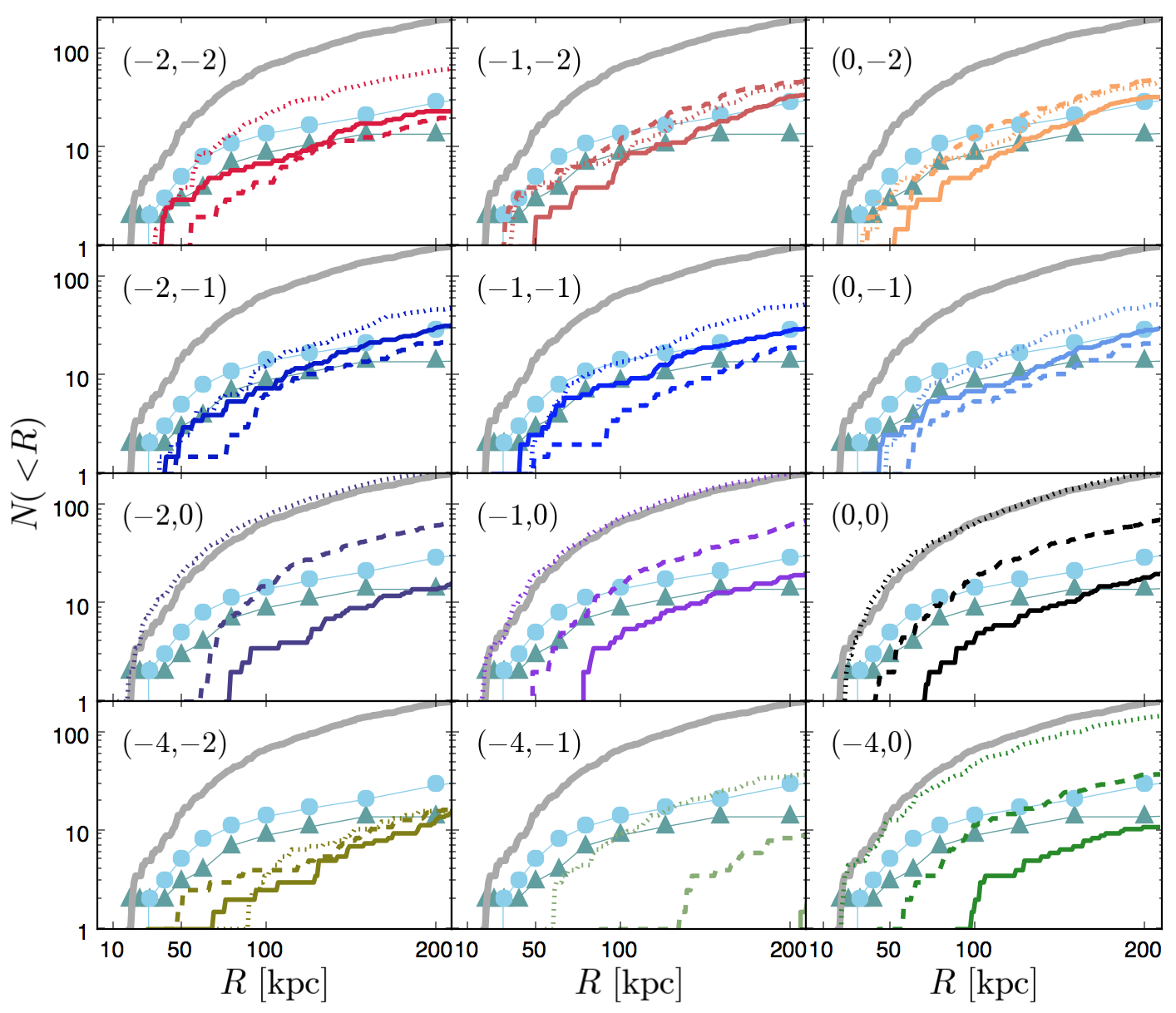}  
\caption[]{\label{fig:RHDP_all} %
Radial subhalo distribution profiles in the 2cDM model with the set of parameters explored and compared. Solid, dashed and dotted curves represent $\sigma_{0}/m = 1, 0.1,$ and $0.01$. 
}
\end{figure*}

In this section, we study the spatial distribution of satellite halos within a MW-type host halo. We first introduce the problem faced by the traditional CDM model in reproducing the radial distribution of such halos centered on the host halo --- the radial halo distribution profile (RHDP) problem. To our knowledge, this has not been previously discussed as one more small-scale CDM problem, akin to SS or TBTF. We then explore the parameter space of the 2cDM model that can alleviate RHDP by comparing our simulation results with the observed MW and M31 satellite distributions.

\subsection{Radial distribution of CDM satellites}

The missing satellite, or SS problem, states that the numerically predicted number of satellite halos surrounding a MW-type host halo greatly exceeds the currently observed satellite counts. The TBTF problem points out that the largest subhalos in simulations have too dense central density to be devoid of observable luminous objects. Not only the larger central density, but also the number of largest subhalos exceeds the observational data. Both of these problems, which were originally motivated by $N$-body numerical simulations and are well presented by the velocity function, imply that the collisionless CDM scenario alone lacks some key ingredients to suppress the number of subhalos and reduce their central mass concentrations. 

Apparently, there can be another small-scale CDM problem, the RHDP problem, which is illustrated  in Figure~\ref{fig:RHDP_CDM}. Here we plot the cumulative number of satellites with radial distance from the center of a host halo, for several simulated CDM halos and the observed MW and M31 halos. In this comparison study, we selected the halos from our simulations, which are large enough, $M_{\rm vir}\ge 10^7M_{\odot}$ and $V_{\rm max} \geq 4$~km~s$^{-1}$, to minimize the effects from numerical artifacts. Correspondingly, the satellite galaxies from observational data were selected based on their total masses of $\gtrsim 10^{7}M_{\odot}$, which is either deduced from kinematics data or estimated given their stellar masses and the mass-to-light ratios (typically at the half-light radius). We should caution that the number of yet undetected dwarfs may still be substantial and can be an increasing function of the radial distance (especially for MW), so introducing observational bias. Understanding this potential problem, we thus set the rather strong and restrictive selection criterion on the dwarfs' mass, which excludes many known ultra-faint dwarfs from consideration. Thus, future detections of ultra-faint dwarfs should not change much the general result. 

In Figure~\ref{fig:RHDP_CDM}, we compare the RHDP of the five most well-resolved halos in the CDM simulations with observations in order to illuminate the discrepancy. The virial masses of the host halos range from $\sim 9 \times 10^{10}M_{\odot}$ to $\sim 5 \times 10^{11}M_{\odot}$, roughly an order of magnitude to a factor of a few smaller than the MW or M31 counterpart, respectively. 
Clearly, only the smallest halo with $M_{\rm vir} \sim 9 \times 10^{10}M_{\odot} \lesssim 0.1 M_\textrm{(MW or M31)}$ shows a reasonably good agreement with the observed MW and M31 satellites at small $R$, but still roughly a factor of a few larger number of subhalos at $R \gtrsim 60 - 80$~kpc. We stress that this halo is only 10\% of the MW by mass. Thus, it seems unlikely that the discrepancy can be fully attributed to the observational bias for the MW-like or Andromeda-like halos.

\subsection{Radial distribution of 2cDM satellites: fiducial model}

The main difficulty the CDM model faces is to reduce the subhalo population in the outer virial range of roughly $R \gtrsim 100$ kpc. Even considering the possibility of still undetected ultra-faint subhalo population far from the halo center, the subhalos with $M_{\rm vir} \gtrsim 10^{7}$M$_{\odot}$ need to be consistently wiped out across the entire virial range to match with the observations. 
In Figure~\ref{fig:RHDP_VSIDM-sig1} we show that 2cDM provides a natural solution to the problem even without the need for baryonic physics. We show the averaged RHDP of the two most well-resolved halos of $M_{\rm vir} \sim 5 \times 10^{11}$ and $\sim 4 \times 10^{11}$M$_{\odot}$ with the same halo selection criteria described above. For comparison purposes, the same fiducial set of parameters are chosen as in Figure~\ref{fig:vmax_VSIDM} for the velocity function.

The most important component of 2cDM being a potential solution to the problem is clearly the mass conversion, which effectively `evaporates' the substructure, thus bringing down the cumulative counts of subhalos to be in rough agreement with observations. This is shown by $(-2,-2)^{\rm conv}_{\rm only}$ and full $(-2,-2)$ (with both mass conversion and elastic scattering) cases sharing a similar trend that shows significant substructure suppressions across the entire virial range as compared to CDM. The scattering-only (i.e.,  SIDM) model, on the other hand, suffers from the excessive number of subhalos akin to the CDM counterpart over the entire range in $R$. We attribute the slightly larger subhalo counts of SIDM compared to CDM to the numerical artifacts due to the relatively large cross-section of $\sigma_{0}/m = 1$. In fact, SIDM with $\sigma_{0}/m = 0.1$ (not shown) follows the CDM curve nearly exactly. We also superpose the two reference cases with the `kick' velocity parameter $V_{k} = 10$ (dotted) and 20 km s$^{-1}$ (dashed) in the figure. 

As shown earlier in Figure~\ref{fig:vmax_VSIDM}, the value of $V_{k}$ roughly determines the break point on the velocity function where the 2cDM model deviates from CDM. The total suppression of the subhalos is therefore stronger for $V_{k} = 20$~km~s$^{-1}$ than that of 10~km~s$^{-1}$. The case with $V_{k} = 20$~km~s$^{-1}$, however, shows that the suppression of subhalos starts around $M_{\rm vir} \lesssim 10^{9}M_{\odot}$, leading to a greater reduction in the cumulative subhalo counts. Yet it still over-predicts the RHDP, which in turn implies the number of subhalos with $M_{\rm vir} \gtrsim 10^{9}M_{\odot}$ need to be substantially reduced. 

Despite that the 2cDM model could reduce the number of subhalos substantially across the virial range compared to the CDM model, we note, again, that the host halos used in Figure~\ref{fig:RHDP_VSIDM-sig1} have a factor of a few smaller masses than the MW or M31 halos. That is, the presented figures underestimate the discrepancy of CDM predictions and simulations by a factor of a few, whereas the 2cDM model remains a plausible solution to the problem. To show this, we extrapolated the cumulative number of subhalos by utilizing the power-law relation in the mass function, $N \propto M^{b}$. In Section~\ref{sec:mf_power_fit}, we found the power-law index for the velocity function to be $b \sim -1$ for CDM, and $-0.8 \lesssim b \lesssim -0.6$ for 2cDM with $0.01 \leq \sigma_{0}/m \leq 1$. With this power-law relation, we can roughly estimate how many more subhalos should be present within the host halo virial radius of a larger mass. For CDM, we find that if the host halo is to become the size of MW or M31 by increasing its virial mass by a factor of $\sim$2, then the cumulative number of halo counts at the lower mass cut-off of $M_{\rm vir} = 10^7M_{\odot}$ would roughly be a factor of three larger. Similarly, for 2cDM, the subhalo counts would be as large as a factor of two at the cut-off mass, roughly doubling the subhalo populations across the virial range. Considering the fact that we have not taken into account the baryonic physics and yet undiscovered dark subhalos, the 2cDM model remains indeed very plausible.

\subsection{Radial distribution of 2cDM satellites: parameter study}

Having shown that our fiducial model is capable of suppressing the low-mass subhalo counts throughout the virial radius region, resulting in good agreement with observations, we now explore the parameter space of the 2cDM model with respect to the RHDP constraint. As is shown earlier for the velocity function, we compare RHDP, computed with the set of 2cDM parameters, with that of the CDM model and observational data in Figure~\ref{fig:RHDP_all}. Each curve represents the averaged RHDP of the two largest halos, same as in Figure~\ref{fig:RHDP_VSIDM-sig1}. The various values of the cross-section values are shown as solid ($\sigma_{0}/m = 1$), dashed (0.1) and dotted (0.01) curves. 

The top two rows explore the six cases of $a_{c} = -2$ and $-1$ with $-2 \leq a_{s} \leq 0$. Given that our 2cDM results are about a factor of two underestimate the overall number of subhalos, all the six cases are within the observationally expected range. For these cases, the difference in the value of the cross-section results in only a a factor of a few difference in the cumulative number of subhalos within the studied virial volume. This implies that the effect of suppressing the subhalos within the virial radius is dominated by the cross-section's velocity-dependence for mass conversion rather than elastic scattering. 

The third row has a subset of cases with no velocity-dependence of the mass conversion cross-section: $(-2,0), (-1,0),$ and $(0,0)$. These are the cases that show the difference among the range of $\sigma_{0}/m$ most clearly, and they are the only 2cDM cases (with the cross-section being as small as $\lesssim 0.01$), which produce RHDPs identical to those of CDM. The general trend is as follows: the larger the cross-section, the stronger the suppression of subhalos so that the curves representing the three different cross-section values are related to each other by simple rescaling. Larger cross-section also seem to result in the strong suppression of subhalos in the inner halo. For example, the closest subhalo to the host halo center can be found at $R \sim 20$ kpc for $\sigma_{0}/m = 0.01$, while it is $R \sim 50$ and 75 kpc for $\sigma_{0}/m = 0.1$ and 1, respectively. Apparently, the structure formation and mergers are slowed down by DM interactions. Furthermore, such a strong suppression of subhalos within 50~kpc of the virial radius is in conflict with observational data and immediately disfavors $\sigma_{0}/m =1$ as the possible candidate model. Doubling the host halo mass to be comparable to the MW halo would increase the subhalo counts by a factor of two as well, which is unlikely to fully resolve the discrepancy. 

The bottom row compares the cases with a cross-section's strong velocity-dependence on the elastic scattering, $a_{s} = -4$. Aside from the fact that these cases are possibly subject to numerical errors rooted in the poorly approximated binary collisions, they tend to over-suppress the subhalo populations, except for $(-4,0)$. While $(-4,0)$ shows a similar trend with the other velocity-independent cases on the mass conversion, its results are not nearly as trustworthy as others since the DM self-interaction occurs in a a strongly-interacting (perhaps, nearly `fluid') regime. It is unclear why $(-4,-1)$ shows a stronger suppression of subhalos compared to $(-4,-2)$. Regardless of this, both $(-4,-1)$ and $(-4,-2)$ are also disfavored models once they are compared with the observational data.

\section{Evolution of matter power spectra}
\label{sec:spectrum}

In Section \ref{sec:vmax_evo} we studied the evolution of the maximum circular velocity function of halos to probe the substructure abundance at a high redshift by comparing the 2cDM models with the CDM model. In this section we extend the study of the substructure abundance by probing the density fluctuations seen in the matter power spectrum. Our simulations made with the small box size of 3$h^{-1}$Mpc side length allow us to probe relatively large $k$ range (a few~$h\,\textrm{Mpc}^{-1}\lesssim k \lesssim 10^{3} h\,\textrm{Mpc}^{-1}$) and help us to study the 2cDM's role in the suppression of the small structure growth and evolution at higher redshifts of $z \sim 2 - 4$. 

In literature there has been numerous studies done on constraining the model parameters for non-traditional DM models other than $\Lambda$CDM by studying the Lyman-$\alpha$ forest flux power spectrum \citep[e.g.][]{narayanan2000, seljak2006, viel2013, wang2014, baur2017}. 
Ly-$\alpha$ forest has been introduced in the seventies by Roger Lynds, who noted a large number of absorption lines in the spectrum of a quasar. Spectral Ly-$\alpha$ transition absorption lines of a neutral hydrogen are produced by electrons transitioning between the ground state ($n=1$) and the first excited state ($n=2$). The Ly-$\alpha$ spectral line has a rest frame wavelength of 1216~$\AA$. The Ly-$\alpha$ absorption lines in the quasar spectra result from intergalactic gas through which the quasar's light has traveled. Since neutral hydrogen clouds in the intergalactic medium are at different redshift, $z$, their absorption lines are observed at different wavelengths. Each individual cloud leaves its fingerprint as an absorption line at a different position in the observed spectrum. It appears that the Ly-$\alpha$ forest becomes a powerful tool for exploring the high-$z$ universe and can greatly constrain non-CDM dark matter models. 

Ideally, hydrodynamical effects must be considered to accurately investigate the Ly-$\alpha$ forest in simulations to reproduce the low-density intergalactic medium since they could enhance the non-linear evolution even at high redshift of $z \sim 2 - 4$ where baryonic feedback effects are not believed to play a significant role. Yet for our purposes to show the suppression of the small scale density fluctuations, here we present the matter power spectrum computed with our $N$-body simulations. We used a publicly available code \citep{bird2017} to compute the power spectrum for GADGET output. 

\begin{figure*}
\centering
\includegraphics[scale = 0.5]{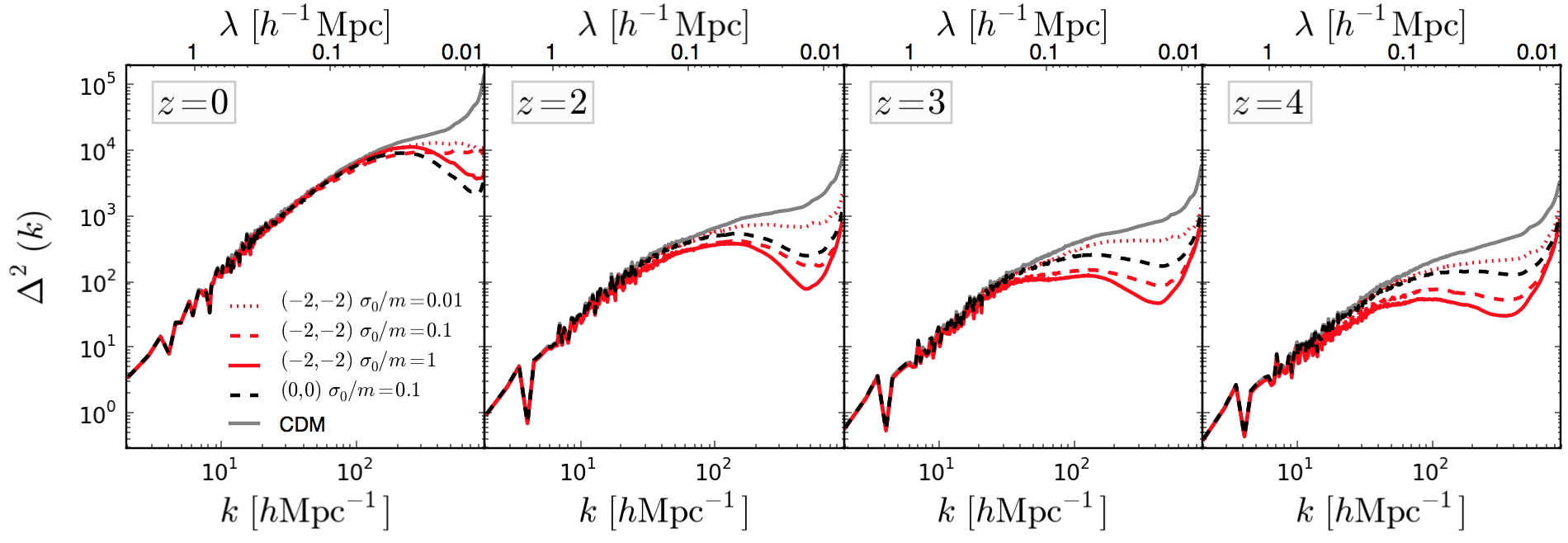}  
\caption[]{\label{fig:Pk_dimless} %
Redshift evolution of the dimensionless power spectra. Only selected cases are shown to elucidate the difference between the 2cDM and CDM models. 
}
\end{figure*}

\begin{figure*}
\centering
\includegraphics[scale = 0.55]{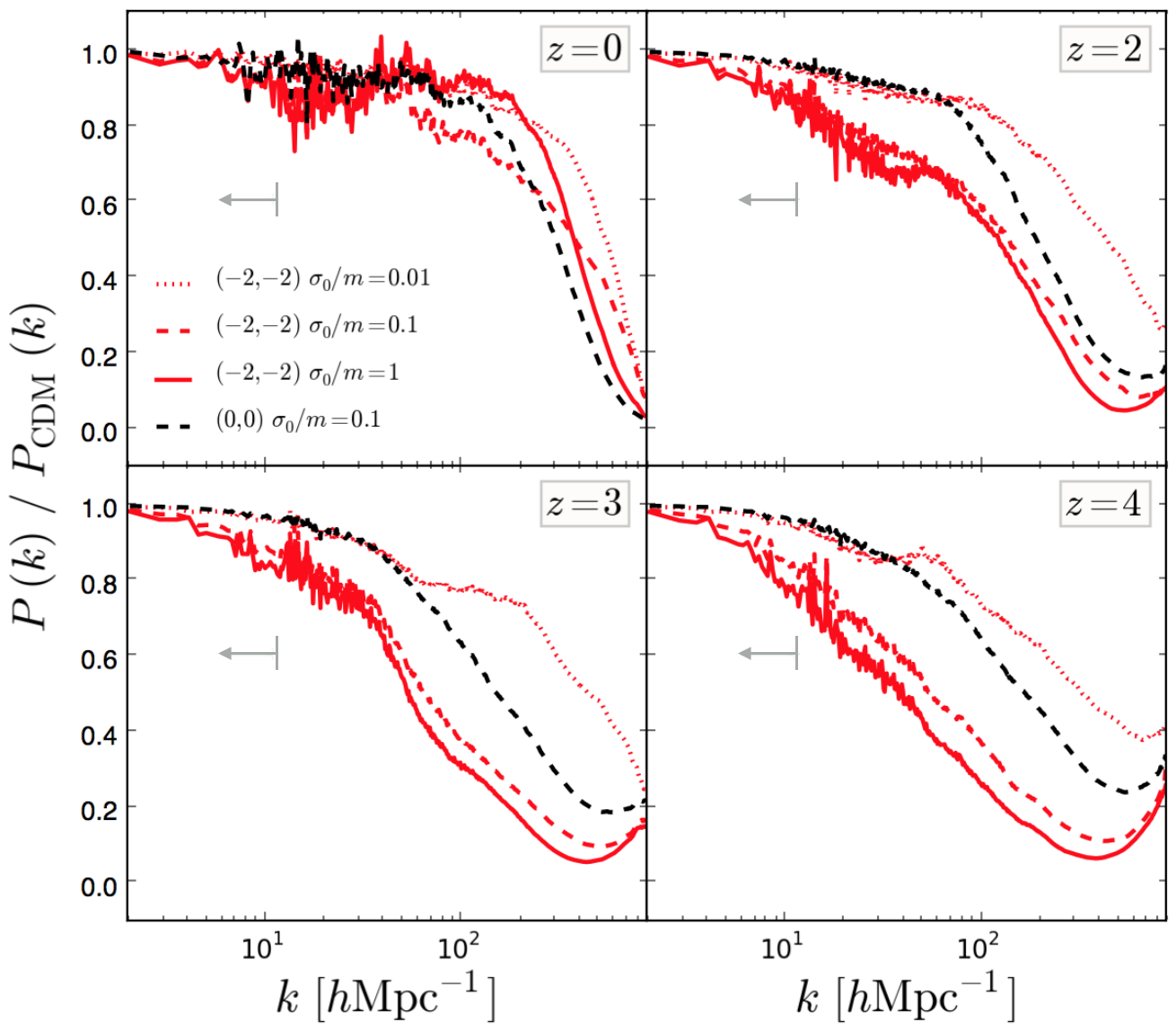}  
\caption[]{\label{fig:Pk_norm} %
Redshift evolution of the normalized power spectra of the 2cDM models to the corresponding CDM model. The upper limit arrow provides an approximated limit in wavenumber range from observational data of MIKE and HIRES shown in \citet{viel2013}.
}
\end{figure*}

Figure \ref{fig:Pk_dimless} and \ref{fig:Pk_norm} show the dimensionless power spectra, $\Delta^{2}(k)=4\pi(2\pi)^{-3}k^3 P(k)$, and the ratio of 2cDM to CDM power spectra, respectively. In both figures, a clear sign of stronger suppression at smaller scales is seen at higher redshift for the 2cDM models. In general the suppression is stronger compared to CDM with either (i) a larger cross-section value, $\sigma_0/m$, or (ii) a stronger power-law index, such as $(a_{s}, a_{c}) = (-2,-2)$, for the velocity-dependent cross-section $\sigma(v)$.
Meanwhile, the deviation from CDM becomes less significant at lower redshift for 2cDM within the explored parameter range. At $z = 0$ the difference of $\sim 10\%$ can only be seen at $k \gtrsim$100 $h$Mpc$^{-1}$. 

To see the difference in the choice of the power-law index for the velocity-dependent cross-section, we also present a $(0,0)$ model with $\sigma_{0}/m = 0.1$ (black dashed curve). It shows distinctively different effects on the suppression scales compared to the $(-2,-2)$ counterpart. That is, with the same $\sigma_{0}/m = 0.1$, $(-2,-2)$ shows a stronger suppression at $k \gtrsim 10$ $h$Mpc$^{-1}$ at $z \geq 2$, while at the largest $k$ range ($\gtrsim 200$ $h$Mpc$^{-1}$) $(0,0)$ overtakes both $(-2,-2)$ with $\sigma_{0}/m = 0.1$ and even $\sigma_{0}/m = 1$. This is clearly due to the increase in the mean DM velocities at lower redshift as the structure formation growth further takes place, and hence the stronger velocity dependence of the cross-section of $(-2,-2)$ (or effectively $\sigma \propto 1/v^{2}$) starts to show a weaker suppression of the substructures compared to $(0,0)$ at lower redshift.

\begin{figure*}
\centering
\includegraphics[scale = 0.8]{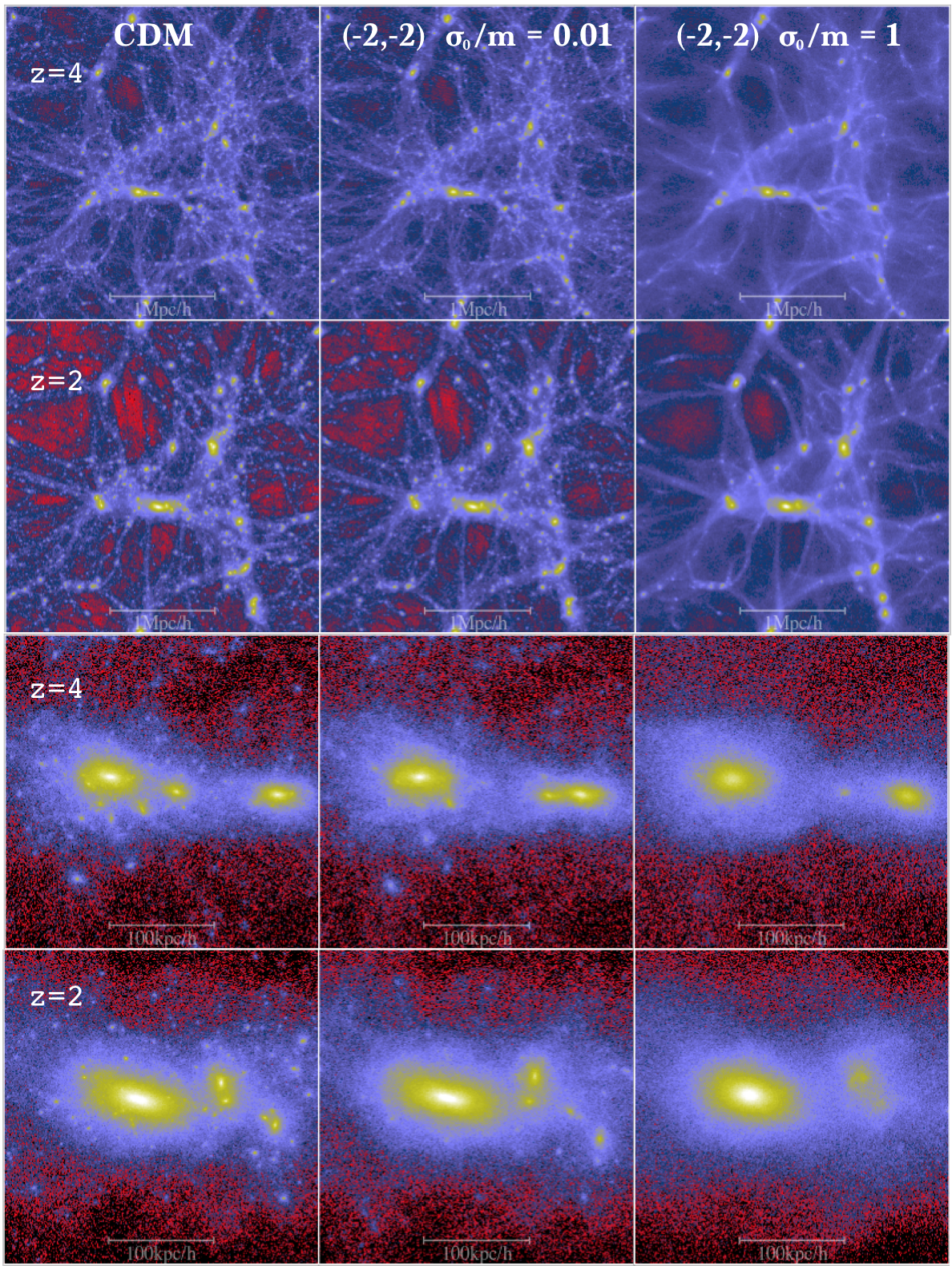} 
\caption[]{\label{fig:viz_highz} %
Dark matter distribution for the 2cDM (right and middle columns) and CDM (left column) models at high redshift. Top two rows show a projected density distribution in a 3$h^{-1}$Mpc side length cubic box, whereas the bottom two rows are for 0.3$h^{-1}$Mpc, focusing on the major halo and its immediate surroundings. 
}
\end{figure*}

In short, the 2cDM model can sufficiently erase the small scale density fluctuations and the magnitude of the deviation from the CDM model depends on the cross-section and the power-law indices for the velocity-dependent cross-section. Figure~\ref{fig:viz_highz} helps one to see it qualitatively. To illustrate, we compare the projected DM density distribution of CDM with the two 2cDM models of $(-2,-2)$ with $\sigma_{0}/m = 0.01$ and 1 at $z=2$ and 4 on 3 Mpc comoving scale (top two rows) and that of 0.3 Mpc (bottom two rows). It shows that on a few Mpc scale the difference in the large scale structure between CDM and $(-2,-2)$ $\sigma_{0}/m = 0.01$ is minimal, while $(-2,-2)$ $\sigma_{0}/m = 1$ has much more smoothed out distribution. 
On the sub-Mpc scale, however, the former two cases start to show some clear differences, especially that the number of substructure has been mildly reduced in $(-2,-2)$ $\sigma_{0}/m = 0.01$. 
The case with $(-2,-2)$ $\sigma_{0}/m = 1$  shows even a stronger suppression, which is strong enough to alter the shape of the inner halo to be rounder compared to that of the triaxial shape in the other two cases.

\section{Summary} \label{sec:CN}

In this paper, we have presented results from an extraordinarily large set of $N$-body cosmological simulations with the simplest two-component (2cDM) model with a very large range of model parameters studied, with halo masses $M \lesssim 10^{12} M_{\odot}$. 
This work aimed primarily at 
(i) exploring the parameter space of the 2cDM model, and 
(ii) setting the constrains on the magnitude and the velocity-dependence of the DM cross-section
through comparison with available observational data in order to deduce the `promising' set of parameters of the 2cDM model.
In our approach, we carefully examine each of the 2cDM parameter set with the maximum circular velocity function to see whether the small-scale cosmological problems rooted in the CDM paradigm, namely the substructure and too-big-to-fail problems, are alleviated. 
Our choice of the parameter set effectively covers four orders of magnitude in the DM cross-section size $\sigma_{0}/m$ (in cm$^2$ g$^{-1}$) from 0.01 to 10, and the DM cross-section's velocity dependence $\sigma \sim v^{a_{s}}$ and $\sim v^{a_{c}}$, or symbolically ($a_{s},a_{c}$), for elastic scattering and inelastic conversion, respectively, going from $a_{s}$ (or $a_{c}$) $= 0$ (i.e., no velocity dependence) to the inverse power-law dependence of $a_{s}$ (or $a_{c}$) as $= -1$, $-2$, and $-4$. 
We performed over sixty simulations with medium resolution to throughly explore the combination of the above parameter set.

The main result of this paper is the confirmation of great success of the 2cDM model in reproducing the suppression of the maximum circular velocity function below a constant critical value. 
The key physical process to the success is the mass conversion, or quantum evaporation, that is unique to 2cDM. 
The model allows DM particles to undergo mass conversions between the two mass states, `heavy' and `light', through inelastic interactions, whereas the mass difference deduced appears to be $\Delta m/m \sim 10^{-7}-10^{-8}$ \citep{medvedev2014}. 
In consequence, the equivalent energy is converted to the kinetic energy of the interacted DM particles, which receive a velocity `kick' $V_{k}$, hence allowing them to escape the halo if $V_{k} > v_{esc}$. Note that elastic scattering alone, as in SIDM, does \emph{not} suppress substructure, leaving the velocity function virtually identical to that of the CDM counterpart. 

The substructure and too-big-to-fail problems are therefore both robustly resolved by 2cDM with a set of available parameters. 
We find that regardless of the DM cross-section's velocity dependence ($a_{s},a_{c}$), the size of the DM cross-section $\sigma_{0}/m$ of 0.1 -- 1 shows good agreement with observations. 
$\sigma_{0}/m = 0.01$ can also be a good candidate, given that in this work we did not consider baryonic effects, especially stellar feedback, which would likely enhance the suppression of the substructures by providing additional sources of energy to disrupt them. Similarly, UV background could also add to the substructure suppression mechanism, which to some extent may alter the DM-only results presented in this work.
We rule out $\sigma_{0}/m = 10$ because it appears to show either an excessive suppression of substructure and/or creation of unnaturally larger halos compared to the CDM counterpart. The exception is the ($-4,-2$) case; however, we take any cases with $a_{s} = -4$ with caution due to its large number of particle interactions that makes DM particles fluid-like, especially at high-$z$, thus invalidating our numerical implementation of the DM-interaction module, based on the assumption of the rare binary collisional interactions.

The study presented here is limited by the lack of baryonic physics, such as gas and stellar dynamics and evolution, ionizing UV radiation, supernovae and black hole feedback and other processes. Their influence on the dark matter distribution is still debated. We plan to explore the role of baryons within the $N$cDM model in the future.


\section*{Acknowledgements}

The simulations for this work were carried out at the Advanced Computing Facilities at the University of Kansas. 
Special thanks to Jay Neitling for kindly allowing us to use the Open Cluster in the Department of Physics and Astronomy at the University of Nevada, Las Vegas. The authors thank Simeon Bird for his help on making the script for computing the power spectrum available to us. We are grateful to Scott Tremaine, Matias Zaldarriaga, Mark Vogelsberger, Lars Hernquist, Ramesh Narayan and Avi Loeb for valuable discussions. MM is grateful to the Institute for Theory and Computation at Harvard University for support and hospitality during his sabbatical leave and acknowledges DOE grant DE-SC0016368.

\bibliographystyle{mnras} 
\bibliography{2cDM-MW-papers12}    

\begin{thebibliography}{}
\makeatletter
\relax
\def\mn@urlcharsother{\let\do\@makeother \do\$\do\&\do\#\do\^\do\_\do\%\do\~}
\def\mn@doi{\begingroup\mn@urlcharsother \@ifnextchar [ {\mn@doi@}
  {\mn@doi@[]}}
\def\mn@doi@[#1]#2{\def\@tempa{#1}\ifx\@tempa\@empty \href
  {http://dx.doi.org/#2} {doi:#2}\else \href {http://dx.doi.org/#2} {#1}\fi
  \endgroup}
\def\mn@eprint#1#2{\mn@eprint@#1:#2::\@nil}
\def\mn@eprint@arXiv#1{\href {http://arxiv.org/abs/#1} {{\tt arXiv:#1}}}
\def\mn@eprint@dblp#1{\href {http://dblp.uni-trier.de/rec/bibtex/#1.xml}
  {dblp:#1}}
\def\mn@eprint@#1:#2:#3:#4\@nil{\def\@tempa {#1}\def\@tempb {#2}\def\@tempc
  {#3}\ifx \@tempc \@empty \let \@tempc \@tempb \let \@tempb \@tempa \fi \ifx
  \@tempb \@empty \def\@tempb {arXiv}\fi \@ifundefined
  {mn@eprint@\@tempb}{\@tempb:\@tempc}{\expandafter \expandafter \csname
  mn@eprint@\@tempb\endcsname \expandafter{\@tempc}}}

\bibitem[\protect\citeauthoryear{{Ahn} \& {Shapiro}}{{Ahn} \&
  {Shapiro}}{2005}]{ahn2005}
{Ahn} K.,  {Shapiro} P.~R.,  2005, \mn@doi [MNRAS]
  {10.1111/j.1365-2966.2005.09492.x}, \href
  {http://adsabs.harvard.edu/abs/2005MNRAS.363.1092A} {363, 1092}

\bibitem[\protect\citeauthoryear{{Baur}, {Palanque-Delabrouille}, {Y{\`e}che},
  {Boyarsky}, {Ruchayskiy}, {Armengaud}  \& {Lesgourgues}}{{Baur}
  et~al.}{2017}]{baur2017}
{Baur} J.,  {Palanque-Delabrouille} N.,  {Y{\`e}che} C.,  {Boyarsky} A.,
  {Ruchayskiy} O.,  {Armengaud} {\'E}.,   {Lesgourgues} J.,  2017, preprint,
  \href {http://adsabs.harvard.edu/abs/2017arXiv170603118B} {} (\mn@eprint
  {arXiv} {1706.03118})

\bibitem[\protect\citeauthoryear{{Bechtol} et~al.,}{{Bechtol}
  et~al.}{2015}]{bechtol2015}
{Bechtol} K.,  et~al., 2015, \mn@doi [\apj] {10.1088/0004-637X/807/1/50}, \href
  {http://adsabs.harvard.edu/abs/2015ApJ...807...50B} {807, 50}

\bibitem[\protect\citeauthoryear{{Belokurov} et~al.,}{{Belokurov}
  et~al.}{2007}]{belokurov2007}
{Belokurov} V.,  et~al., 2007, \mn@doi [\apj] {10.1086/509718}, \href
  {http://adsabs.harvard.edu/abs/2007ApJ...654..897B} {654, 897}

\bibitem[\protect\citeauthoryear{{Bird}}{{Bird}}{2017}]{bird2017}
{Bird} S.,  2017, {GenPK: Power spectrum generator}, Astrophysics Source Code
  Library (\mn@eprint {ascl} {1706.006})

\bibitem[\protect\citeauthoryear{{Blumenthal}, {Faber}, {Flores}  \&
  {Primack}}{{Blumenthal} et~al.}{1986}]{blumenthal1986}
{Blumenthal} G.~R.,  {Faber} S.~M.,  {Flores} R.,   {Primack} J.~R.,  1986,
  \mn@doi [\apj] {10.1086/163867}, \href
  {http://adsabs.harvard.edu/abs/1986ApJ...301...27B} {301, 27}

\bibitem[\protect\citeauthoryear{{Boylan-Kolchin}, {Bullock}  \&
  {Kaplinghat}}{{Boylan-Kolchin} et~al.}{2011}]{boylan-kolchin2011}
{Boylan-Kolchin} M.,  {Bullock} J.~S.,   {Kaplinghat} M.,  2011, \mn@doi
  [MNRAS] {10.1111/j.1745-3933.2011.01074.x}, \href
  {http://adsabs.harvard.edu/abs/2011MNRAS.415L..40B} {415, L40}

\bibitem[\protect\citeauthoryear{{Bramante}, {Fox}, {Kribs}  \&
  {Martin}}{{Bramante} et~al.}{2016}]{bramante2016}
{Bramante} J.,  {Fox} P.~J.,  {Kribs} G.~D.,   {Martin} A.,  2016, preprint,
  \href {http://adsabs.harvard.edu/abs/2016arXiv160802662B} {} (\mn@eprint
  {arXiv} {1608.02662})

\bibitem[\protect\citeauthoryear{{Brook} \& {Di Cintio}}{{Brook} \& {Di
  Cintio}}{2015}]{brook2015}
{Brook} C.~B.,  {Di Cintio} A.,  2015, \mn@doi [\mnras] {10.1093/mnras/stv864},
  \href {http://adsabs.harvard.edu/abs/2015MNRAS.450.3920B} {450, 3920}

\bibitem[\protect\citeauthoryear{{Col{\'{\i}}n}, {Avila-Reese}, {Valenzuela}
  \& {Firmani}}{{Col{\'{\i}}n} et~al.}{2002}]{colin2002}
{Col{\'{\i}}n} P.,  {Avila-Reese} V.,  {Valenzuela} O.,   {Firmani} C.,  2002,
  \mn@doi [\apj] {10.1086/344259}, \href
  {http://adsabs.harvard.edu/abs/2002ApJ...581..777C} {581, 777}

\bibitem[\protect\citeauthoryear{{Collett} et~al.,}{{Collett}
  et~al.}{2017}]{collett2017}
{Collett} T.~E.,  et~al., 2017, preprint, \href
  {http://adsabs.harvard.edu/abs/2017arXiv170308410C} {} (\mn@eprint {arXiv}
  {1703.08410})

\bibitem[\protect\citeauthoryear{{Diemand}, {Kuhlen}, {Madau}, {Zemp}, {Moore},
  {Potter}  \& {Stadel}}{{Diemand} et~al.}{2008}]{diemand2008}
{Diemand} J.,  {Kuhlen} M.,  {Madau} P.,  {Zemp} M.,  {Moore} B.,  {Potter} D.,
    {Stadel} J.,  2008, \mn@doi [\nat] {10.1038/nature07153}, \href
  {http://adsabs.harvard.edu/abs/2008Natur.454..735D} {454, 735}

\bibitem[\protect\citeauthoryear{{Efstathiou}}{{Efstathiou}}{1992}]{efstathiou1992}
{Efstathiou} G.,  1992, \mn@doi [\mnras] {10.1093/mnras/256.1.43P}, \href
  {http://adsabs.harvard.edu/abs/1992MNRAS.256P..43E} {256, 43P}

\bibitem[\protect\citeauthoryear{{Errani}, {Pe{\~n}arrubia}, {Laporte}  \&
  {G{\'o}mez}}{{Errani} et~al.}{2017}]{errani2017}
{Errani} R.,  {Pe{\~n}arrubia} J.,  {Laporte} C.~F.~P.,   {G{\'o}mez} F.~A.,
  2017, \mn@doi [\mnras] {10.1093/mnrasl/slw211}, \href
  {http://adsabs.harvard.edu/abs/2017MNRAS.465L..59E} {465, L59}

\bibitem[\protect\citeauthoryear{{Flores} \& {Primack}}{{Flores} \&
  {Primack}}{1994}]{flores1994}
{Flores} R.~A.,  {Primack} J.~R.,  1994, \mn@doi [\apjl] {10.1086/187350},
  \href {http://adsabs.harvard.edu/abs/1994ApJ...427L...1F} {427, L1}

\bibitem[\protect\citeauthoryear{{Fry} et~al.,}{{Fry} et~al.}{2015}]{fry2015}
{Fry} A.~B.,  et~al., 2015, \mn@doi [\mnras] {10.1093/mnras/stv1330}, \href
  {http://adsabs.harvard.edu/abs/2015MNRAS.452.1468F} {452, 1468}

\bibitem[\protect\citeauthoryear{{Garrison-Kimmel}, {Rocha}, {Boylan-Kolchin},
  {Bullock}  \& {Lally}}{{Garrison-Kimmel} et~al.}{2013}]{garrison-kimmel2013}
{Garrison-Kimmel} S.,  {Rocha} M.,  {Boylan-Kolchin} M.,  {Bullock} J.~S.,
  {Lally} J.,  2013, \mn@doi [\mnras] {10.1093/mnras/stt984}, \href
  {http://adsabs.harvard.edu/abs/2013MNRAS.433.3539G} {433, 3539}

\bibitem[\protect\citeauthoryear{{Garrison-Kimmel}, {Boylan-Kolchin}, {Bullock}
   \& {Kirby}}{{Garrison-Kimmel} et~al.}{2014}]{garrison-kimmel2014}
{Garrison-Kimmel} S.,  {Boylan-Kolchin} M.,  {Bullock} J.~S.,   {Kirby} E.~N.,
  2014, \mn@doi [\mnras] {10.1093/mnras/stu1477}, \href
  {http://adsabs.harvard.edu/abs/2014MNRAS.444..222G} {444, 222}

\bibitem[\protect\citeauthoryear{{Governato} et~al.,}{{Governato}
  et~al.}{2012}]{governato2012}
{Governato} F.,  et~al., 2012, \mn@doi [MNRAS]
  {10.1111/j.1365-2966.2012.20696.x}, \href
  {http://adsabs.harvard.edu/abs/2012MNRAS.422.1231G} {422, 1231}

\bibitem[\protect\citeauthoryear{{Graham}, {Harnik}, {Rajendran}  \&
  {Saraswat}}{{Graham} et~al.}{2010}]{graham2010}
{Graham} P.~W.,  {Harnik} R.,  {Rajendran} S.,   {Saraswat} P.,  2010, \mn@doi
  [\prd] {10.1103/PhysRevD.82.063512}, \href
  {http://adsabs.harvard.edu/abs/2010PhRvD..82f3512G} {82, 063512}

\bibitem[\protect\citeauthoryear{{Hambrick}, {Ostriker}, {Johansson}  \&
  {Naab}}{{Hambrick} et~al.}{2011}]{hambrick2011}
{Hambrick} D.~C.,  {Ostriker} J.~P.,  {Johansson} P.~H.,   {Naab} T.,  2011,
  \mn@doi [\mnras] {10.1111/j.1365-2966.2011.18312.x}, \href
  {http://adsabs.harvard.edu/abs/2011MNRAS.413.2421H} {413, 2421}

\bibitem[\protect\citeauthoryear{{Hayashi}, {Navarro}, {Taylor}, {Stadel}  \&
  {Quinn}}{{Hayashi} et~al.}{2003}]{hayashi2003}
{Hayashi} E.,  {Navarro} J.~F.,  {Taylor} J.~E.,  {Stadel} J.,   {Quinn} T.,
  2003, \mn@doi [\apj] {10.1086/345788}, \href
  {http://adsabs.harvard.edu/abs/2003ApJ...584..541H} {584, 541}

\bibitem[\protect\citeauthoryear{{Hobbs}, {Read}, {Agertz}, {Iannuzzi}  \&
  {Power}}{{Hobbs} et~al.}{2016}]{hobbs2016}
{Hobbs} A.,  {Read} J.~I.,  {Agertz} O.,  {Iannuzzi} F.,   {Power} C.,  2016,
  \mn@doi [\mnras] {10.1093/mnras/stw251}, \href
  {http://adsabs.harvard.edu/abs/2016MNRAS.458..468H} {458, 468}

\bibitem[\protect\citeauthoryear{{Hu}, {Barkana}  \& {Gruzinov}}{{Hu}
  et~al.}{2000}]{hu2000}
{Hu} W.,  {Barkana} R.,   {Gruzinov} A.,  2000, \mn@doi [Physical Review
  Letters] {10.1103/PhysRevLett.85.1158}, \href
  {http://adsabs.harvard.edu/abs/2000PhRvL..85.1158H} {85, 1158}

\bibitem[\protect\citeauthoryear{{Hui}, {Ostriker}, {Tremaine}  \&
  {Witten}}{{Hui} et~al.}{2017}]{hui2017}
{Hui} L.,  {Ostriker} J.~P.,  {Tremaine} S.,   {Witten} E.,  2017, \mn@doi
  [\prd] {10.1103/PhysRevD.95.043541}, \href
  {http://adsabs.harvard.edu/abs/2017PhRvD..95d3541H} {95, 043541}

\bibitem[\protect\citeauthoryear{{Kamada}, {Kaplinghat}, {Pace}  \&
  {Yu}}{{Kamada} et~al.}{2017}]{kamada+17}
{Kamada} A.,  {Kaplinghat} M.,  {Pace} A.~B.,   {Yu} H.-B.,  2017, \mn@doi
  [Physical Review Letters] {10.1103/PhysRevLett.119.111102}, \href
  {http://adsabs.harvard.edu/abs/2017PhRvL.119k1102K} {119, 111102}

\bibitem[\protect\citeauthoryear{{Kazantzidis}, {Mayer}, {Mastropietro},
  {Diemand}, {Stadel}  \& {Moore}}{{Kazantzidis}
  et~al.}{2004}]{kazantzidis2004}
{Kazantzidis} S.,  {Mayer} L.,  {Mastropietro} C.,  {Diemand} J.,  {Stadel} J.,
    {Moore} B.,  2004, \mn@doi [\apj] {10.1086/420840}, \href
  {http://adsabs.harvard.edu/abs/2004ApJ...608..663K} {608, 663}

\bibitem[\protect\citeauthoryear{{Klypin}, {Kravtsov}, {Valenzuela}  \&
  {Prada}}{{Klypin} et~al.}{1999}]{klypin1999}
{Klypin} A.,  {Kravtsov} A.~V.,  {Valenzuela} O.,   {Prada} F.,  1999, \mn@doi
  [\apj] {10.1086/307643}, \href
  {http://adsabs.harvard.edu/abs/1999ApJ...522...82K} {522, 82}

\bibitem[\protect\citeauthoryear{{Klypin}, {Kravtsov}, {Bullock}  \&
  {Primack}}{{Klypin} et~al.}{2001}]{klypin2001}
{Klypin} A.,  {Kravtsov} A.~V.,  {Bullock} J.~S.,   {Primack} J.~R.,  2001,
  \mn@doi [\apj] {10.1086/321400}, \href
  {http://adsabs.harvard.edu/abs/2001ApJ...554..903K} {554, 903}

\bibitem[\protect\citeauthoryear{{Klypin}, {Karachentsev}, {Makarov}  \&
  {Nasonova}}{{Klypin} et~al.}{2015}]{klypin2015}
{Klypin} A.,  {Karachentsev} I.,  {Makarov} D.,   {Nasonova} O.,  2015, \mn@doi
  [\mnras] {10.1093/mnras/stv2040}, \href
  {http://adsabs.harvard.edu/abs/2015MNRAS.454.1798K} {454, 1798}

\bibitem[\protect\citeauthoryear{{Knollmann} \& {Knebe}}{{Knollmann} \&
  {Knebe}}{2009}]{knollmann2009}
{Knollmann} S.~R.,  {Knebe} A.,  2009, \mn@doi [\apjs]
  {10.1088/0067-0049/182/2/608}, \href
  {http://adsabs.harvard.edu/abs/2009ApJS..182..608K} {182, 608}

\bibitem[\protect\citeauthoryear{{Kormendy} \& {Freeman}}{{Kormendy} \&
  {Freeman}}{2016}]{kormendy2016}
{Kormendy} J.,  {Freeman} K.~C.,  2016, \mn@doi [\apj]
  {10.3847/0004-637X/817/2/84}, \href
  {http://adsabs.harvard.edu/abs/2016ApJ...817...84K} {817, 84}

\bibitem[\protect\citeauthoryear{{Kuflik}, {Perelstein}, {Lorier}  \&
  {Tsai}}{{Kuflik} et~al.}{2016}]{kuflik2016}
{Kuflik} E.,  {Perelstein} M.,  {Lorier} N.~R.-L.,   {Tsai} Y.-D.,  2016,
  \mn@doi [Physical Review Letters] {10.1103/PhysRevLett.116.221302}, \href
  {http://adsabs.harvard.edu/abs/2016PhRvL.116v1302K} {116, 221302}

\bibitem[\protect\citeauthoryear{{Laevens} et~al.,}{{Laevens}
  et~al.}{2015}]{laevens2015}
{Laevens} B.~P.~M.,  et~al., 2015, \mn@doi [\apj] {10.1088/0004-637X/813/1/44},
  \href {http://adsabs.harvard.edu/abs/2015ApJ...813...44L} {813, 44}

\bibitem[\protect\citeauthoryear{{Markevitch}, {Gonzalez}, {Clowe},
  {Vikhlinin}, {Forman}, {Jones}, {Murray}  \& {Tucker}}{{Markevitch}
  et~al.}{2004}]{markevitch2004}
{Markevitch} M.,  {Gonzalez} A.~H.,  {Clowe} D.,  {Vikhlinin} A.,  {Forman} W.,
   {Jones} C.,  {Murray} S.,   {Tucker} W.,  2004, \mn@doi [\apj]
  {10.1086/383178}, \href {http://adsabs.harvard.edu/abs/2004ApJ...606..819M}
  {606, 819}

\bibitem[\protect\citeauthoryear{{Martizzi}, {Teyssier}, {Moore}  \&
  {Wentz}}{{Martizzi} et~al.}{2012}]{martizzi2012}
{Martizzi} D.,  {Teyssier} R.,  {Moore} B.,   {Wentz} T.,  2012, \mn@doi
  [\mnras] {10.1111/j.1365-2966.2012.20879.x}, \href
  {http://adsabs.harvard.edu/abs/2012MNRAS.422.3081M} {422, 3081}

\bibitem[\protect\citeauthoryear{{Martizzi}, {Teyssier}  \& {Moore}}{{Martizzi}
  et~al.}{2013}]{martizzi2013}
{Martizzi} D.,  {Teyssier} R.,   {Moore} B.,  2013, \mn@doi [\mnras]
  {10.1093/mnras/stt297}, \href
  {http://adsabs.harvard.edu/abs/2013MNRAS.432.1947M} {432, 1947}

\bibitem[\protect\citeauthoryear{{Mateo}}{{Mateo}}{1998}]{mateo1998}
{Mateo} M.~L.,  1998, \mn@doi [\araa] {10.1146/annurev.astro.36.1.435}, \href
  {http://adsabs.harvard.edu/abs/1998ARA%26A..36..435M} {36, 435}

\bibitem[\protect\citeauthoryear{{McCullough} \& {Randall}}{{McCullough} \&
  {Randall}}{2013}]{mccullough2013}
{McCullough} M.,  {Randall} L.,  2013, \mn@doi [\jcap]
  {10.1088/1475-7516/2013/10/058}, \href
  {http://adsabs.harvard.edu/abs/2013JCAP...10..058M} {10, 058}

\bibitem[\protect\citeauthoryear{{Medvedev}}{{Medvedev}}{2010a}]{medvedev2010}
{Medvedev} M.~V.,  2010a, preprint, \href
  {http://adsabs.harvard.edu/abs/2010arXiv1004.3377M} {} (\mn@eprint {arXiv}
  {1004.3377})

\bibitem[\protect\citeauthoryear{{Medvedev}}{{Medvedev}}{2010b}]{medvedev2010b}
{Medvedev} M.~V.,  2010b, \mn@doi [Journal of Physics A Mathematical General]
  {10.1088/1751-8113/43/37/372002}, \href
  {http://adsabs.harvard.edu/abs/2010JPhA...43K2002M} {43, 372002}

\bibitem[\protect\citeauthoryear{{Medvedev}}{{Medvedev}}{2014a}]{medvedev2014theo}
{Medvedev} M.~V.,  2014a, \mn@doi [\jcap] {10.1088/1475-7516/2014/06/063},
  \href {http://adsabs.harvard.edu/abs/2014JCAP...06..063M} {6, 063}

\bibitem[\protect\citeauthoryear{{Medvedev}}{{Medvedev}}{2014b}]{medvedev2014}
{Medvedev} M.~V.,  2014b, \mn@doi [Physical Review Letters]
  {10.1103/PhysRevLett.113.071303}, \href
  {http://adsabs.harvard.edu/abs/2014PhRvL.113g1303M} {113, 071303}

\bibitem[\protect\citeauthoryear{{Moore}, {Ghigna}, {Governato}, {Lake},
  {Quinn}, {Stadel}  \& {Tozzi}}{{Moore} et~al.}{1999}]{moore1999}
{Moore} B.,  {Ghigna} S.,  {Governato} F.,  {Lake} G.,  {Quinn} T.,  {Stadel}
  J.,   {Tozzi} P.,  1999, \mn@doi [\apjl] {10.1086/312287}, \href
  {http://adsabs.harvard.edu/abs/1999ApJ...524L..19M} {524, L19}

\bibitem[\protect\citeauthoryear{{Moore}, {Gelato}, {Jenkins}, {Pearce}  \&
  {Quilis}}{{Moore} et~al.}{2000}]{moore2000}
{Moore} B.,  {Gelato} S.,  {Jenkins} A.,  {Pearce} F.~R.,   {Quilis} V.,  2000,
  \mn@doi [ApJL] {10.1086/312692}, \href
  {http://adsabs.harvard.edu/abs/2000ApJ...535L..21M} {535, L21}

\bibitem[\protect\citeauthoryear{{Narayanan}, {Spergel}, {Dav{\'e}}  \&
  {Ma}}{{Narayanan} et~al.}{2000}]{narayanan2000}
{Narayanan} V.~K.,  {Spergel} D.~N.,  {Dav{\'e}} R.,   {Ma} C.-P.,  2000,
  \mn@doi [\apjl] {10.1086/317269}, \href
  {http://adsabs.harvard.edu/abs/2000ApJ...543L.103N} {543, L103}

\bibitem[\protect\citeauthoryear{{Navarro}, {Frenk}  \& {White}}{{Navarro}
  et~al.}{1996}]{navarro1996}
{Navarro} J.~F.,  {Frenk} C.~S.,   {White} S.~D.~M.,  1996, \mn@doi [\apj]
  {10.1086/177173}, \href {http://adsabs.harvard.edu/abs/1996ApJ...462..563N}
  {462, 563}

\bibitem[\protect\citeauthoryear{{Newman}, {Treu}, {Ellis}  \& {Sand}}{{Newman}
  et~al.}{2013}]{newman2013b}
{Newman} A.~B.,  {Treu} T.,  {Ellis} R.~S.,   {Sand} D.~J.,  2013, \mn@doi
  [\apj] {10.1088/0004-637X/765/1/25}, \href
  {http://adsabs.harvard.edu/abs/2013ApJ...765...25N} {765, 25}

\bibitem[\protect\citeauthoryear{{O{\~n}orbe}, {Boylan-Kolchin}, {Bullock},
  {Hopkins}, {Kere{\v s}}, {Faucher-Gigu{\`e}re}, {Quataert}  \&
  {Murray}}{{O{\~n}orbe} et~al.}{2015}]{Onorbe2015}
{O{\~n}orbe} J.,  {Boylan-Kolchin} M.,  {Bullock} J.~S.,  {Hopkins} P.~F.,
  {Kere{\v s}} D.,  {Faucher-Gigu{\`e}re} C.-A.,  {Quataert} E.,   {Murray} N.,
   2015, \mn@doi [\mnras] {10.1093/mnras/stv2072}, \href
  {http://adsabs.harvard.edu/abs/2015MNRAS.454.2092O} {454, 2092}

\bibitem[\protect\citeauthoryear{{Okamoto}, {Gao}  \& {Theuns}}{{Okamoto}
  et~al.}{2008}]{okamoto2008}
{Okamoto} T.,  {Gao} L.,   {Theuns} T.,  2008, \mn@doi [\mnras]
  {10.1111/j.1365-2966.2008.13830.x}, \href
  {http://adsabs.harvard.edu/abs/2008MNRAS.390..920O} {390, 920}

\bibitem[\protect\citeauthoryear{{Papastergis} \& {Shankar}}{{Papastergis} \&
  {Shankar}}{2016}]{papastergis2016}
{Papastergis} E.,  {Shankar} F.,  2016, \mn@doi [\aap]
  {10.1051/0004-6361/201527854}, \href
  {http://adsabs.harvard.edu/abs/2016A%26A...591A..58P} {591, A58}

\bibitem[\protect\citeauthoryear{{Papastergis}, {Martin}, {Giovanelli}  \&
  {Haynes}}{{Papastergis} et~al.}{2011}]{papastergis2011}
{Papastergis} E.,  {Martin} A.~M.,  {Giovanelli} R.,   {Haynes} M.~P.,  2011,
  \mn@doi [\apj] {10.1088/0004-637X/739/1/38}, \href
  {http://adsabs.harvard.edu/abs/2011ApJ...739...38P} {739, 38}

\bibitem[\protect\citeauthoryear{{Papastergis}, {Giovanelli}, {Haynes}  \&
  {Shankar}}{{Papastergis} et~al.}{2015}]{papastergis2015}
{Papastergis} E.,  {Giovanelli} R.,  {Haynes} M.~P.,   {Shankar} F.,  2015,
  \mn@doi [\aap] {10.1051/0004-6361/201424909}, \href
  {http://adsabs.harvard.edu/abs/2015A%26A...574A.113P} {574, A113}

\bibitem[\protect\citeauthoryear{{Pe{\~n}arrubia}, {Pontzen}, {Walker}  \&
  {Koposov}}{{Pe{\~n}arrubia} et~al.}{2012}]{penarrubia2012}
{Pe{\~n}arrubia} J.,  {Pontzen} A.,  {Walker} M.~G.,   {Koposov} S.~E.,  2012,
  \mn@doi [\apjl] {10.1088/2041-8205/759/2/L42}, \href
  {http://adsabs.harvard.edu/abs/2012ApJ...759L..42P} {759, L42}

\bibitem[\protect\citeauthoryear{{Planck Collaboration} et~al.,}{{Planck
  Collaboration} et~al.}{2015}]{planck2015}
{Planck Collaboration} et~al., 2015, preprint, \href
  {http://adsabs.harvard.edu/abs/2015arXiv150201589P} {} (\mn@eprint {arXiv}
  {1502.01589})

\bibitem[\protect\citeauthoryear{{Pontzen} \& {Governato}}{{Pontzen} \&
  {Governato}}{2012}]{pontzen2012}
{Pontzen} A.,  {Governato} F.,  2012, \mn@doi [\mnras]
  {10.1111/j.1365-2966.2012.20571.x}, \href
  {http://adsabs.harvard.edu/abs/2012MNRAS.421.3464P} {421, 3464}

\bibitem[\protect\citeauthoryear{{Power}, {Robotham}, {Obreschkow}, {Hobbs}  \&
  {Lewis}}{{Power} et~al.}{2016}]{power2016}
{Power} C.,  {Robotham} A.~S.~G.,  {Obreschkow} D.,  {Hobbs} A.,   {Lewis}
  G.~F.,  2016, preprint, \href
  {http://adsabs.harvard.edu/abs/2016arXiv160602038P} {} (\mn@eprint {arXiv}
  {1606.02038})

\bibitem[\protect\citeauthoryear{{Quinn}, {Katz}  \& {Efstathiou}}{{Quinn}
  et~al.}{1996}]{quinn1996}
{Quinn} T.,  {Katz} N.,   {Efstathiou} G.,  1996, \mn@doi [\mnras]
  {10.1093/mnras/278.4.L49}, \href
  {http://adsabs.harvard.edu/abs/1996MNRAS.278L..49Q} {278, L49}

\bibitem[\protect\citeauthoryear{{Randall}, {Markevitch}, {Clowe}, {Gonzalez}
  \& {Brada{\v c}}}{{Randall} et~al.}{2008}]{randall2008}
{Randall} S.~W.,  {Markevitch} M.,  {Clowe} D.,  {Gonzalez} A.~H.,   {Brada{\v
  c}} M.,  2008, \mn@doi [\apj] {10.1086/587859}, \href
  {http://adsabs.harvard.edu/abs/2008ApJ...679.1173R} {679, 1173}

\bibitem[\protect\citeauthoryear{{Richardson} et~al.,}{{Richardson}
  et~al.}{2011}]{richardson2011}
{Richardson} J.~C.,  et~al., 2011, \mn@doi [\apj] {10.1088/0004-637X/732/2/76},
  \href {http://adsabs.harvard.edu/abs/2011ApJ...732...76R} {732, 76}

\bibitem[\protect\citeauthoryear{{Rocha}, {Peter}, {Bullock}, {Kaplinghat},
  {Garrison-Kimmel}, {O{\~n}orbe}  \& {Moustakas}}{{Rocha}
  et~al.}{2013}]{rocha2013}
{Rocha} M.,  {Peter} A.~H.~G.,  {Bullock} J.~S.,  {Kaplinghat} M.,
  {Garrison-Kimmel} S.,  {O{\~n}orbe} J.,   {Moustakas} L.~A.,  2013, \mn@doi
  [MNRAS] {10.1093/mnras/sts514}, \href
  {http://adsabs.harvard.edu/abs/2013MNRAS.430...81R} {430, 81}

\bibitem[\protect\citeauthoryear{{Seljak}, {Makarov}, {McDonald}  \&
  {Trac}}{{Seljak} et~al.}{2006}]{seljak2006}
{Seljak} U.,  {Makarov} A.,  {McDonald} P.,   {Trac} H.,  2006, \mn@doi
  [Physical Review Letters] {10.1103/PhysRevLett.97.191303}, \href
  {http://adsabs.harvard.edu/abs/2006PhRvL..97s1303S} {97, 191303}

\bibitem[\protect\citeauthoryear{{Simon} \& {Geha}}{{Simon} \&
  {Geha}}{2007}]{simon2007}
{Simon} J.~D.,  {Geha} M.,  2007, \mn@doi [\apj] {10.1086/521816}, \href
  {http://adsabs.harvard.edu/abs/2007ApJ...670..313S} {670, 313}

\bibitem[\protect\citeauthoryear{{Simon} et~al.,}{{Simon}
  et~al.}{2011}]{simon2011}
{Simon} J.~D.,  et~al., 2011, \mn@doi [\apj] {10.1088/0004-637X/733/1/46},
  \href {http://adsabs.harvard.edu/abs/2011ApJ...733...46S} {733, 46}

\bibitem[\protect\citeauthoryear{{Somerville}}{{Somerville}}{2002}]{summerville2002}
{Somerville} R.~S.,  2002, \mn@doi [\apjl] {10.1086/341444}, \href
  {http://adsabs.harvard.edu/abs/2002ApJ...572L..23S} {572, L23}

\bibitem[\protect\citeauthoryear{{Spergel} \& {Steinhardt}}{{Spergel} \&
  {Steinhardt}}{2000}]{spergel2000}
{Spergel} D.~N.,  {Steinhardt} P.~J.,  2000, \mn@doi [Physical Review Letters]
  {10.1103/PhysRevLett.84.3760}, \href
  {http://adsabs.harvard.edu/abs/2000PhRvL..84.3760S} {84, 3760}

\bibitem[\protect\citeauthoryear{{Springel}}{{Springel}}{2005}]{springel2005}
{Springel} V.,  2005, \mn@doi [\mnras] {10.1111/j.1365-2966.2005.09655.x},
  \href {http://adsabs.harvard.edu/abs/2005MNRAS.364.1105S} {364, 1105}

\bibitem[\protect\citeauthoryear{{Springel} et~al.,}{{Springel}
  et~al.}{2008}]{springel2008}
{Springel} V.,  et~al., 2008, \mn@doi [\mnras]
  {10.1111/j.1365-2966.2008.14066.x}, \href
  {http://adsabs.harvard.edu/abs/2008MNRAS.391.1685S} {391, 1685}

\bibitem[\protect\citeauthoryear{{Swaters}, {Sancisi}, {van Albada}  \& {van
  der Hulst}}{{Swaters} et~al.}{2009}]{swaters2009}
{Swaters} R.~A.,  {Sancisi} R.,  {van Albada} T.~S.,   {van der Hulst} J.~M.,
  2009, \mn@doi [\aap] {10.1051/0004-6361:200810516}, \href
  {http://adsabs.harvard.edu/abs/2009A%26A...493..871S} {493, 871}

\bibitem[\protect\citeauthoryear{{Turner}}{{Turner}}{1983}]{turner1983}
{Turner} M.~S.,  1983, \mn@doi [\prd] {10.1103/PhysRevD.28.1243}, \href
  {http://adsabs.harvard.edu/abs/1983PhRvD..28.1243T} {28, 1243}

\bibitem[\protect\citeauthoryear{{Viel}, {Becker}, {Bolton}  \&
  {Haehnelt}}{{Viel} et~al.}{2013}]{viel2013}
{Viel} M.,  {Becker} G.~D.,  {Bolton} J.~S.,   {Haehnelt} M.~G.,  2013, \mn@doi
  [\prd] {10.1103/PhysRevD.88.043502}, \href
  {http://adsabs.harvard.edu/abs/2013PhRvD..88d3502V} {88, 043502}

\bibitem[\protect\citeauthoryear{{Vogelsberger} \& {Zavala}}{{Vogelsberger} \&
  {Zavala}}{2013}]{vogelsberger2013}
{Vogelsberger} M.,  {Zavala} J.,  2013, \mn@doi [\mnras]
  {10.1093/mnras/sts712}, \href
  {http://adsabs.harvard.edu/abs/2013MNRAS.430.1722V} {430, 1722}

\bibitem[\protect\citeauthoryear{{Vogelsberger}, {Zavala}, {Simpson}  \&
  {Jenkins}}{{Vogelsberger} et~al.}{2014}]{vogelsberger2014}
{Vogelsberger} M.,  {Zavala} J.,  {Simpson} C.,   {Jenkins} A.,  2014, \mn@doi
  [MNRAS] {10.1093/mnras/stu1713}, \href
  {http://adsabs.harvard.edu/abs/2014MNRAS.444.3684V} {444, 3684}

\bibitem[\protect\citeauthoryear{{Vogelsberger}, {Zavala}, {Cyr-Racine},
  {Pfrommer}, {Bringmann}  \& {Sigurdson}}{{Vogelsberger}
  et~al.}{2016}]{vogelsberger2016}
{Vogelsberger} M.,  {Zavala} J.,  {Cyr-Racine} F.-Y.,  {Pfrommer} C.,
  {Bringmann} T.,   {Sigurdson} K.,  2016, \mn@doi [\mnras]
  {10.1093/mnras/stw1076}, \href
  {http://adsabs.harvard.edu/abs/2016MNRAS.460.1399V} {460, 1399}

\bibitem[\protect\citeauthoryear{{Wang}, {Peter}, {Strigari}, {Zentner},
  {Arant}, {Garrison-Kimmel}  \& {Rocha}}{{Wang} et~al.}{2014}]{wang2014}
{Wang} M.-Y.,  {Peter} A.~H.~G.,  {Strigari} L.~E.,  {Zentner} A.~R.,  {Arant}
  B.,  {Garrison-Kimmel} S.,   {Rocha} M.,  2014, \mn@doi [\mnras]
  {10.1093/mnras/stu1747}, \href
  {http://adsabs.harvard.edu/abs/2014MNRAS.445..614W} {445, 614}

\bibitem[\protect\citeauthoryear{{Wetzel}, {Hopkins}, {Kim},
  {Faucher-Gigu{\`e}re}, {Kere{\v s}}  \& {Quataert}}{{Wetzel}
  et~al.}{2016}]{wetzel2016}
{Wetzel} A.~R.,  {Hopkins} P.~F.,  {Kim} J.-h.,  {Faucher-Gigu{\`e}re} C.-A.,
  {Kere{\v s}} D.,   {Quataert} E.,  2016, \mn@doi [\apjl]
  {10.3847/2041-8205/827/2/L23}, \href
  {http://adsabs.harvard.edu/abs/2016ApJ...827L..23W} {827, L23}

\bibitem[\protect\citeauthoryear{{Willman} et~al.,}{{Willman}
  et~al.}{2005}]{willman2005}
{Willman} B.,  et~al., 2005, \mn@doi [\apjl] {10.1086/431760}, \href
  {http://adsabs.harvard.edu/abs/2005ApJ...626L..85W} {626, L85}

\bibitem[\protect\citeauthoryear{{Yoshida}, {Springel}, {White}  \&
  {Tormen}}{{Yoshida} et~al.}{2000}]{yoshida2000}
{Yoshida} N.,  {Springel} V.,  {White} S.~D.~M.,   {Tormen} G.,  2000, \mn@doi
  [\apjl] {10.1086/317306}, \href
  {http://adsabs.harvard.edu/abs/2000ApJ...544L..87Y} {544, L87}

\bibitem[\protect\citeauthoryear{{Zavala}, {Jing}, {Faltenbacher}, {Yepes},
  {Hoffman}, {Gottl{\"o}ber}  \& {Catinella}}{{Zavala}
  et~al.}{2009}]{zavala2009}
{Zavala} J.,  {Jing} Y.~P.,  {Faltenbacher} A.,  {Yepes} G.,  {Hoffman} Y.,
  {Gottl{\"o}ber} S.,   {Catinella} B.,  2009, \mn@doi [\apj]
  {10.1088/0004-637X/700/2/1779}, \href
  {http://adsabs.harvard.edu/abs/2009ApJ...700.1779Z} {700, 1779}

\bibitem[\protect\citeauthoryear{{de Blok} \& {Bosma}}{{de Blok} \&
  {Bosma}}{2002}]{deblok2002}
{de Blok} W.~J.~G.,  {Bosma} A.,  2002, \mn@doi [\aap]
  {10.1051/0004-6361:20020080}, \href
  {http://adsabs.harvard.edu/abs/2002A%26A...385..816D} {385, 816}

\bibitem[\protect\citeauthoryear{{de Blok}, {McGaugh}, {Bosma}  \& {Rubin}}{{de
  Blok} et~al.}{2001}]{deblok2001}
{de Blok} W.~J.~G.,  {McGaugh} S.~S.,  {Bosma} A.,   {Rubin} V.~C.,  2001,
  \mn@doi [\apjl] {10.1086/320262}, \href
  {http://adsabs.harvard.edu/abs/2001ApJ...552L..23D} {552, L23}

\makeatother
\end{thebibliography}

\appendix


\section{Pair-wise Interaction Probability}

\begin{figure*}
\centering
\includegraphics[scale = 0.6]{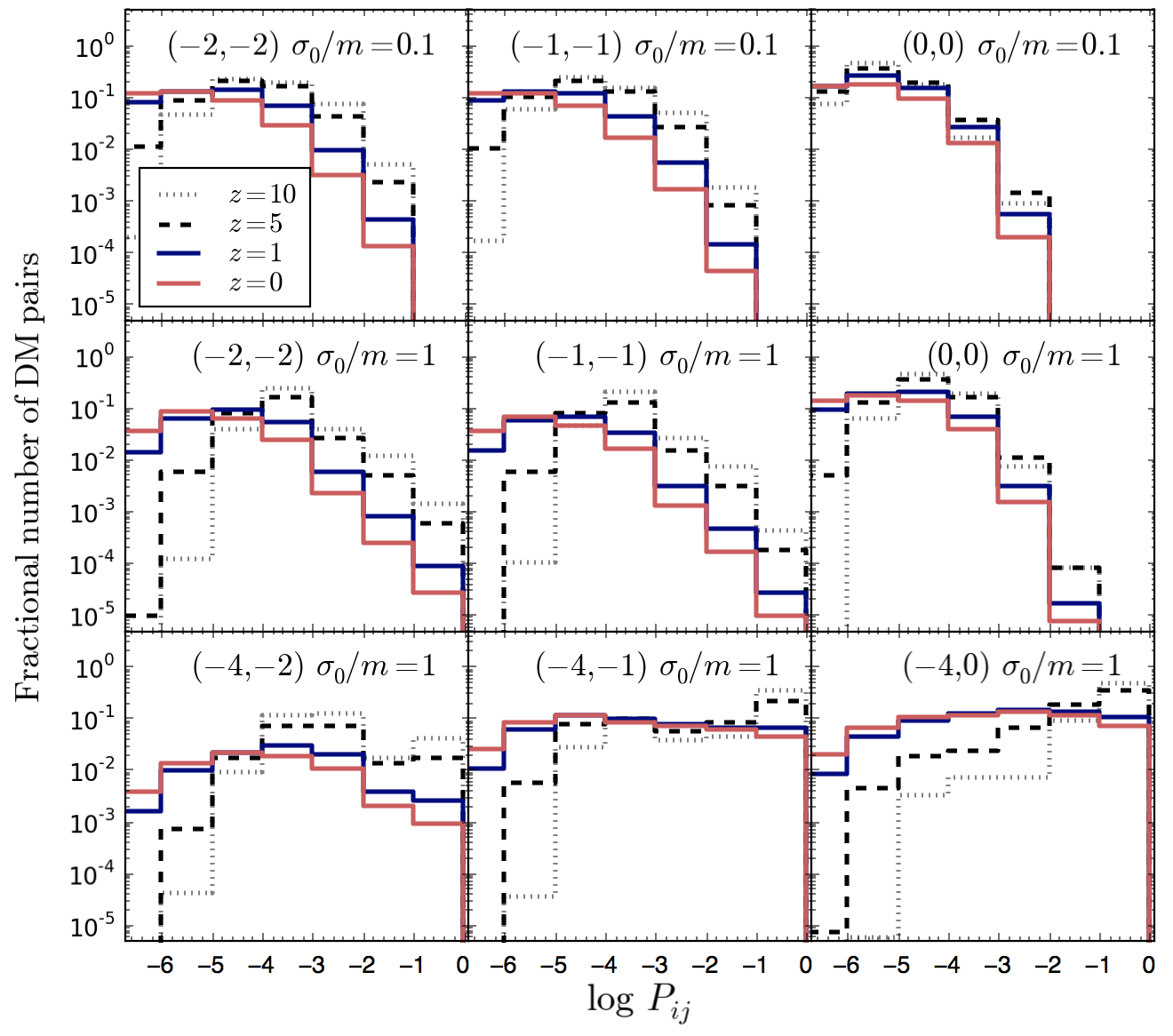}  
\caption[]{\label{fig:int_prob} %
Fractional number of DM pairs as a function of the DM-pair interaction probabilities. 
The number of DM-pairs is counted cumulatively since the beginning of the simulation, and the number of pairs in each probability bin is normalized by the cumulative total number of DM-pairs formed. Probability bins smaller than $\sim10^{-7}$ are omitted for simplicity. 
}
\end{figure*}

In the 2cDM model, we assume a set of power-law indices for the velocity-dependent cross-section.
In principle, a strong velocity-dependence of the DM cross-section with negative powers can cause large interaction probabilities among DM pairs when their velocities are small, especially in the early Universe.  
Such a large interaction probability could render the rare binary collision approximation used in the 2cDM model invalid, and put it in the fluid regime. 
To check the validity of the assumption, we examined the interaction probabilities $P_{ij}$ for all the DM pairs formed at different redshifts.
Figure~\ref{fig:int_prob} presents the fractional number of DM pairs, which is the normalized cumulative number of DM pairs for each interaction probability (Eq.~(\ref{eq:probability})) bin, at redshifts of $z=0,1,5,$ and 10.
We chose some of the extreme cases to scrutinize the validity of the approximation.

\begin{figure}
\centering
\includegraphics[scale = 0.5]{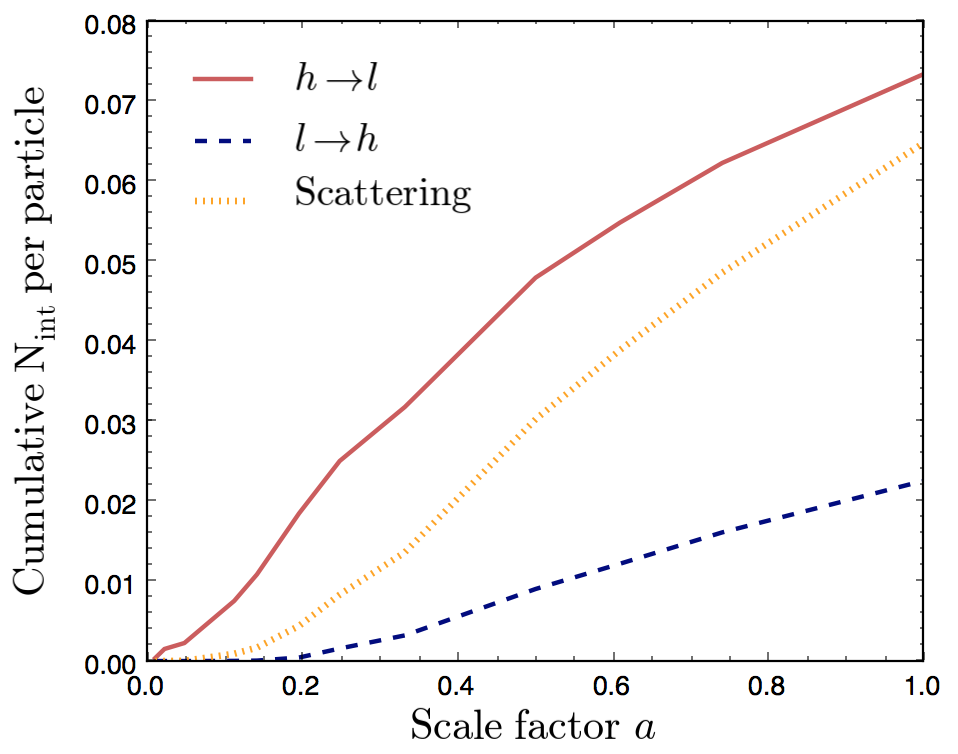}  
\caption[]{\label{fig:N_interactions} %
Cumulative number of interactions per particle versus the scale factor for a 2cDM model. A particular model of $(0,0)$ with $\sigma_{0}/m = 0.1$ was used.
}
\end{figure}

The top and middle rows show the symmetric cases ($a_{s} = a_{c}$) with $\sigma_{0}/m = 0.1$ and 1, respectively. As is shown, the fractional number of DM pairs can be well below 10$^{-4}$ for $P_{ij} > 0.1$ at $z=0$ for those cases.  
At higher $z$, however, the cases with $a_{s}, a_{c} < 0$ consistently show an increase in the fractional number of DM pairs in the larger $P_{ij}$ bins. This is due to the fact that the negative power-law dependence of the DM cross-section in the earlier Universe causing an increase in the interaction probabilities for DM pairs with smaller velocities.
For a reference, a DM pair at $5 < z < 10$ can have $P_{ij}$ roughly more than an oder of magnitude larger than that of at $z=0$ in the range of $10^{-4} \leq P_{ij}{\rm \ (bin)} \leq 1$. 
Even so, the figure shows the fractional number of DM pairs in $P_{ij} > 0.1$ bin can only be as large as 10$^{-3}$, i.e., one in a thousand DM pairs, at $z=10$ for ($-2,-2$) with $\sigma_{0}/m = 1$. 
We also explored the asymmetric cases of ($a_{s} \neq a_{c}$) and found the results are similar.

Overall, our results show that the rare binary collision approximation is not compromised and remains valid over the cosmic time scale for the models with $-2 \leq a_{s}, a_{c} \leq 0$ and $0.01 \leq \sigma_{0}/m \leq 1$. The goodness of the approximation improves at later times when DM pairs experience self-interactions most actively. We conducted the resolution test on this and presented separately in the Appendix.

As the most extreme cases we studied in this work, the bottom row shows the $a_{s} = -4$ series with $\sigma_{0}/m = 1$. Not surprisingly, the number of pairs that have $P_{ij} > 0.1$ is considerably larger than that of the cases with $-2 \leq a_{s}, a_{c} \leq 0$ presented in the top and middle rows. 
Such a steep velocity dependence on the DM cross-section enhances the number of DM interactions dramatically in the early Universe. At $z=0$, both $(-4,-1)$ and $(-4,0)$ cases show $\sim5\%$ and $\sim8\%$, respectively, of the DM pairs having $P_{ij} > 0.1$. Their respective values can be raised to $\sim36\%$ and $52\%$ at $z=10$. In this regime DM can be more accurately depicted as `fluid', and hence our rare binary approximation becomes inaccurate.

Lastly, Figure \ref{fig:N_interactions} shows the cumulative number of interactions per particle as a function of scale factor through DM pair-wise interactions that result in either mass conversions ({\it heavy} to {\it light}, $h \rightarrow l$ or {\it light} to {\it heavy}, $l \rightarrow h$) or elastic scattering. 
For this study we chose a case with $(0,0)$ $\sigma_{0}/m = 0.1$. The figure shows the conversion process of $h \rightarrow l$ remains to be the dominant mass conversion mechanism since the heavy ones tend to remain inside halos even after they interacted so that they have more chances for re-interactions. In contrast, the conversion rate for $l \rightarrow h$ remains sub-dominant over the entire cosmic time scale since the $l$ particles carry larger kinetic energy after they interacted, which in most cases exceed the escape velocities of DM halos in our relatively small box size, thus offering them smaller chances of re-interactions. 
In the meantime, the elastic scattering process continues to increase roughly linearly at $a > 0.2$ (or $z < 4$) which is the direct consequence of the ongoing large structure growth.


\section{Convergence test}
\begin{figure}
  \centering
  \includegraphics[scale = 0.43]{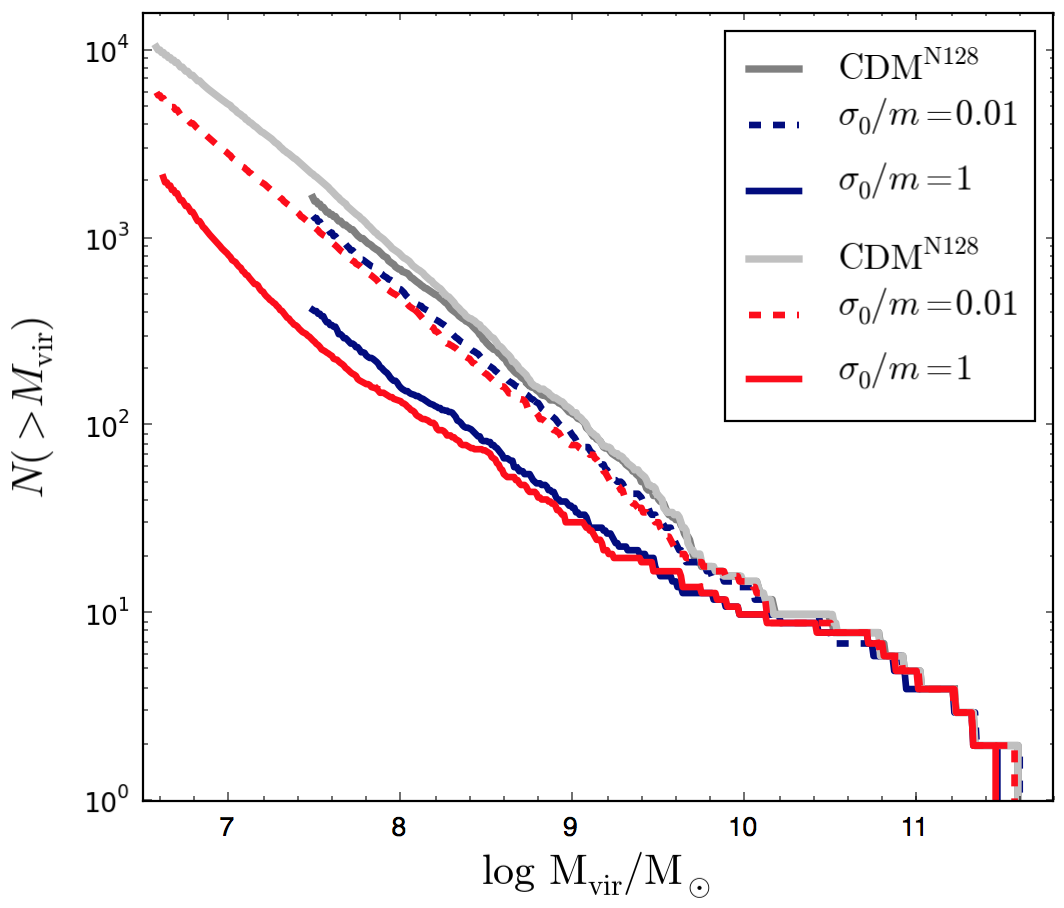}  
  \caption[]{\label{fig:convergence_mass-func} %
Convergence test -- halo mass function: $N=128^3$ (blue) and $256^3$ (red). 2cDM with two different cross-section sizes of $\sigma_{0}/m = 0.01$ and 1 are being compared with CDM. 
  }
\end{figure}

We compared two cases of the total number of particles in the simulation box, $N = 128^{3}$ and $256^{3}$, to check our model's dependency on resolution. For the model in comparison, we chose our fiducial case of $(-2,-2)$. Figure~\ref{fig:convergence_mass-func} compares the two cases with the mass function, the cumulative number of halo counts as a function of halo mass. 
Our 2cDM model appears to be somewhat more prone to the resolution difference at the lower-mass end, though the discrepancy does not alter the general shape. Those low-mass halos identified by the AHF could contain the total number of particles as small as $\lesssim 100$ after applying the halo selection criteria described earlier. It is therefore imperative not to take the face value at the low-mass end. 
Note that the halo counts is slightly reduced at low-mass end in higher resolution than in lower resolution.
This effect is more profound with a larger $\sigma_{0}/m$ value. On the contrary, CDM shows the opposite trend -- a higher resolution producing slightly larger number of low-mass halos. 
This is clearly attributed to the fact that a higher resolution slightly increases the number of DM interactions in 2cDM, resulting in an enhancement in the suppression of the low-mass halos that are not generally well-resolved.

\begin{figure}
  \centering
  \includegraphics[scale = 0.435]{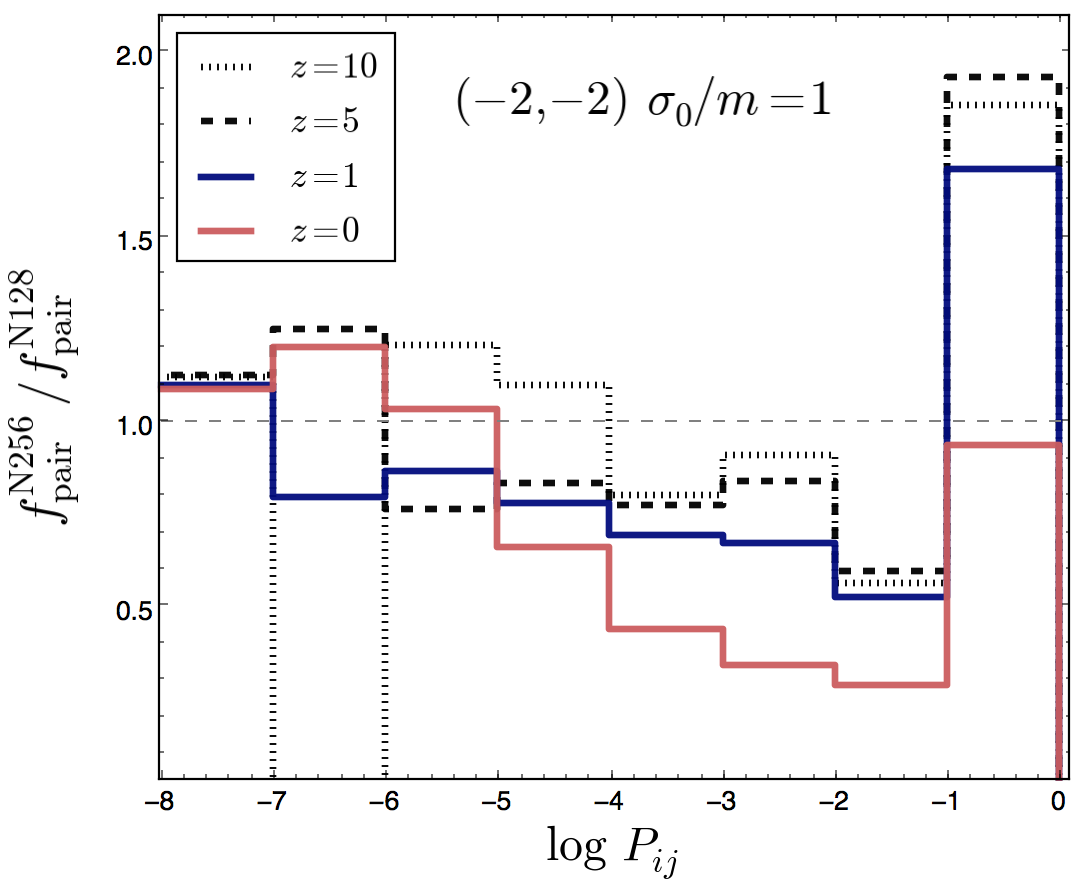}  
  \caption[]{\label{fig:res-test-frac-prob} %
Convergence test -- DM-pair interaction probabilities. As mentioned earlier, a higher resolution increases the number of DM interactions in general, but the difference is kept minimal given the fact that fractional number of DM pairs in the $P_{ij} > 0.1$ is on the order of $10^{-5}$ at $z=0$ and $10^{-3}$ at $z=10$ for $(-2,-2)$ with $\sigma_{0}/m = 1$. 
  }
\end{figure}

In Figure~\ref{fig:res-test-frac-prob} we show another convergence test on the fractional number of DM pairs as a function of the interaction probabilities. 
To see the difference between the two resolutions, the figure shows the ratio of the fractional number of DM pairs, $f^{\rm N256}_{\rm pair} / f^{\rm N128}_{\rm pair}$. 
Our primary interest is the largest probability bin, $P_{ij} > 0.1$. 
A factor of 2 difference seen at higher $z$ in the specific bin is hardly enough to make a significant difference in the results since both $f^{\rm N256}_{\rm pair}$ and $f^{\rm N128}_{\rm pair}$ are on the order of $10^{-5}$ (refer to Figure~\ref{fig:int_prob}). 
Across 8 orders of magnitude in $P_{ij}$, our model shows an excellent convergence in terms of the DM interaction probabilities. Since the case used for this convergence test has a rather large cross-section of $\sigma_{0}/m = 1$, we expect similarly good or even better convergence for the cases with smaller cross-section values.

\label{lastpage}

\end{document}